\documentclass[useAMS,usenatbib]{mn2e}

\usepackage{graphicx}
\usepackage{times}
\usepackage{natbib}
\usepackage{rotating}

\newcommand{\be}{\begin{equation}}
\newcommand{\ee}{\end{equation}}
\newcommand{\ba}{\begin{eqnarray}}
\newcommand{\ea}{\end{eqnarray}}
\newcommand{\brr}{\begin{array}}
\newcommand{\err}{\end{array}}
\newcommand{\bc}{\begin{center}}
\newcommand{\ec}{\end{center}}
\newcommand{\hm}{\,h^{-1}{\rm Mpc}}

\newcommand{\msun}{\,h^{-1}M_\odot}

\newcommand{\vel}{\,{\rm km\,s^{-1}}}

\newcommand{\mincir}{\raise
  -2.truept\hbox{\rlap{\hbox{$\sim$}}\raise5.truept \hbox{$<$}\ }}
\newcommand{\magcir}{\raise
  -2.truept\hbox{\rlap{\hbox{$\sim$}}\raise5.truept \hbox{$>$}\ }}
\newcommand{\siml}{\raise
  -2.truept\hbox{\rlap{\hbox{$\sim$}}\raise5.truept \hbox{$<$}\ }}
\newcommand{\simg}{\raise
  -2.truept\hbox{\rlap{\hbox{$\sim$}}\raise5.truept \hbox{$>$}\ }}

\title[Substructures in hydrodynamical cluster simulations]
{Substructures in hydrodynamical cluster simulations}
\author[K. Dolag, S. Borgani, G. Murante, V. Springel]
{K.~Dolag$^{1}$\thanks{E-mail: kdolag@mpa-garching.mpg.de}, S.
Borgani$^{2,3,4}$, G. Murante$^5$ and V. Springel$^1$\\
$^1$ Max-Planck-Institut f\"ur Astrophysik, Karl-Schwarzschild Strasse
  1, Garching bei M\"unchen, Germany (kdolag@mpa-garching.mpg.de)\\
$^2$ Dipartimento di Astronomia dell'Universit\`a di Trieste, via
  Tiepolo 11, I-34131 Trieste, Italy (borgani@oats.inaf.it)\\
$^3$ INFN -- Istituto Nazionale di Fisica Nucleare, Trieste, Italy\\
$^4$ INAF -- Istituto Nazionale di Astrofisica, Trieste, Italy\\
$^5$ INAF -- Astronomical Observatory of Torino, Str. Osservatorio
25,  I-10025, Pino Torinese, Torino, Italy\\
}

\begin{document}

\date{Accepted ???. Received ???; in original form ???}

\pagerange{\pageref{firstpage}--\pageref{lastpage}} \pubyear{0000}

\maketitle

\label{firstpage}

\begin{abstract}
   The abundance and structure of dark matter subhalos has been
   analyzed extensively in recent studies of dark matter-only
   simulations, but comparatively little is known about the impact of
   baryonic physics on halo substructures. We here extend the {\small
     SUBFIND} algorithm for substructure identification such that it
   can be reliably applied to dissipative hydrodynamical simulations
   that include star formation. This allows, in particular, the
   identification of galaxies as substructures in simulations of
   clusters of galaxies, and a determination of their content of
   gravitationally bound stars, dark matter, and hot and cold
   gas. Using a large set of cosmological cluster simulations, we
   present a detailed analysis of halo substructures in hydrodynamical
   simulations of galaxy clusters, focusing in particular on the
   influence both of radiative and non-radiative gas physics, and of
   non-standard physics such as thermal conduction and feedback by
   galactic outflows. We also examine the impact of numerical nuisance
   parameters such as artificial viscosity parameterizations.  We find
   that diffuse hot gas is efficiently stripped from subhalos when
   they enter the highly pressurized cluster atmosphere. This has the
   effect of
   decreasing the subhalo mass
   function relative to a corresponding dark matter-only
   simulation. These effects are mitigated in radiative runs, where
   baryons condense in the central subhalo regions and form compact
   stellar cores. However, in all cases, only a very small fraction,
   of the order of one percent, of subhalos within the cluster virial
   radii preserve a gravitationally bound hot gaseous atmosphere. The
   fraction of mass contributed by gas in subhalos is found to
   increase with the cluster-centric distance. Interestingly, this
   trend extends well beyond the virial radii, thus showing that
   galaxies feel the environment of the pressurized cluster gas over
   fairly large distances. The compact stellar cores (i.e.~galaxies)
   are generally more resistant against tidal disruption than pure
   dark matter subhalos. Still, the fraction of star-dominated
   substructures within our simulated clusters is only $\sim 10$ per
   cent. We expect that the finite resolution in our simulations makes
   the galaxies overly susceptible to tidal disruption, hence the
   above fraction of star--dominated galaxies should represent a lower
   limit for the actual fraction of galaxies surviving the disruption
   of their host dark matter subhalo.
\end{abstract}



\section{Introduction} \label{sec:intro}

The hierarchical cold dark matter (CDM) model has been established as the
standard model of cosmic structure formation and its parameters are now
tightly constrained by a variety of observations \citep[e.g.][\ \ and references
therein]{2008arXiv0803.0547K, 2006Natur.440.1137S}. Within the CDM scenario,
galaxy clusters are the largest gravitationally bound objects in the universe,
and they exhibit rich information about the process of structure formation.
As a result, they attract great interest both from observational and
theoretical points of view. Due to their recent formation, clusters of
galaxies also represent the ideal places where the hierarchical assembly of
structures is `caught in the act'. Indeed, the complexity of their internal
structure reflects the infall of hundreds, if not thousands, of smaller
objects and their subsequent destruction or survival within the cluster
potential. The dynamical processes determining the fate of the accreting
structures also provide the link between the internal dynamics of clusters and
the properties of their galaxy population.

Modern cosmological simulations provide an ideal tool to study the
non-linear dynamics of these processes in a cosmological
environment.  Thanks to the high resolution reached by dark matter
(DM) only simulations, a consistent picture of the population of
subhalos within galaxy clusters has now emerged \citep[e.g.,
][]{1999ApJ...524L..19M,2000ApJ...544..616G,2001MNRAS.328..726S,2002MNRAS.335L..84S,
2003MNRAS.345.1313S,2004MNRAS.352..535D,2004MNRAS.352L...1G,2004MNRAS.348..333D,2004ApJ...609...35K}.
Although these simulations give a highly detailed description of
the evolution of DM substructures, they can not provide
information on how gas-dynamical processes affect the properties
and the dynamics of substructures. Yet such an understanding is
needed to make a more direct link of the properties of dark matter
substructures with those of the galaxies that orbit in clusters.
For example, we would like to know the timescale over which
subhalos can retain their diffuse gas once they infall into a
cluster, as this also determines the time interval over which gas
can continue to cool and to refuel ongoing star formation.  Being
highly pressurized, the intra--cluster medium (ICM) is quite
efficient in stripping gas from merging structures
(\citealt{1972ApJ...176....1G}; see also
\citealt{2008MNRAS.383..593M} and references therein), thereby
altering both the mass of the subhalos and their survival time.
The stripping also governs the timescale over which galaxy
morphology and colours are modified inside galaxy clusters
\citep{1999MNRAS.308..947A,2004AJ....127.3361K}.  Finally,
although baryons are expected to only play a sub-dominant role for
the dynamics of substructures due to their limited mass fraction,
they may nevertheless alter the structure of subhalos in important
ways, affecting their survival, mass loss rate, abundance and
radial distribution. Quantifying these differences relative to
dark matter-only simulations is an important challenge for the
present generation of cosmological hydrodynamical simulations.

So far there have only been a limited number of investigations of the
properties of sub-halos within hydrodynamical simulations.  Based on
cluster simulationsw with the adaptive mesh refinement code ART,
\citet{2005ApJ...618..557N} have demonstrated that the stellar mass of
infalling galaxies is approximately conserved and the survival of the
galaxies is slightly enhanced. Similar studies using the Tree-SPH code
GASOLINE have been performed by \citet{2006MNRAS.366.1529M}, where
twice as many sub-halos within the central part of a Galaxy sized halo
were found when compared with pure dark matter simulations. Finally
\citet{2008ApJ...678....6W} used a TreeSPH code to investigate the
halo occupation and sub-halo correlation functions for one group-sized
halo, also pointing out the enhanced survival of sub-structures in the
hydrodynamical runs.

Here we present a systematic analysis of a large sample of galaxy
clusters, with an emphasis on the role of non-standard physics and the
details of feedback prescriptions on the survival of
galaxies. Additionally, we also present an analysis of the evolution
of the diffuse gas within the galaxies. In general our findings are in
broad agreement with those in previous studies, but we stress that our
star dominant systems are still not as numerous as expected from
semi-analytic modelling to reproduce the cluster luminosity functions
(e.g., \citet{2007MNRAS.375....2D}; see also
\citet{2006RPPh...69.3101B} for a review on semi-analytic models of
galaxy formation).

Thanks to improvements in numerical codes and higher computer
performance, cosmological hydrodynamical simulations of galaxy
clusters can now model a number of the most important physical
processes in galaxy formation, such as radiative cooling, star
formation, and feedback in energy and metals, while at the same time
reaching sufficient resolution to treat these processes in a
numerically robust and physically meaningful way \citep[e.g., ][\ ,
for a recent review]{2008SSRv..134..269B}. This allows more realistic
structure formation calculations of the coupled system of dark matter
and baryonic gas than possible with dark matter only simulations, far
into the highly non-linear regime.  For instance, radiative cooling
has the effect of letting a significant fraction of the baryons cool
at the halo centres, which alters the shape and concentration of halos
through the mechanism of adiabatic contraction
\citep[e.g.,][]{2004ApJ...616...16G}. At the same time, the
condensation of cooled baryons and their subsequent conversion into
collisionless stars is expected to largely protect them from
stripping, reducing the rate of disruption of substructures and
increasing their survival times. Indeed, assuming that galaxies at
least survive for a while the disruption of their host DM halos has
been shown to be crucial for semi--analytical models of galaxy
formation to provide a successful description of the observed galaxy
population in clusters \citep[e.g., ][]{2004MNRAS.352L...1G}.

This discussion highlights the importance of understanding the properties and
the evolution of cluster substructures as predicted by modern cosmological
hydrodynamical simulations. Evidently, an important technical prerequisite
for such an analysis is the availability of suitable numerical algorithms to
reliably identify substructure in dissipative simulations. To this end we
introduce a modified version of the {\small SUBFIND} algorithm, and test its
robustness.  We then apply it in a detailed analysis of the properties of
substructures in a large ensemble of galaxy clusters, which have been
simulated both at different resolutions in their DM-only version, and in
several further versions where different physical processes where included,
such as non-radiative and radiative hydrodynamics, star formation, energy
feedback, gas viscosity and thermal conduction. The aim of this analysis is to
quantify the numerical robustness of the measured properties of substructures,
and to identify the quantitative influence of these physical processes on
substructure statistics in comparison with dark matter only models.

The paper is organized as follows. We describe in Section~\ref{sec:sim} the
sample of galaxy clusters and the simulations performed.  Section~\ref{sec:detection} provides a description of the algorithm used to identify
gravitationally bound substructures. This algorithm is a modified version of
SUBFIND \citep{2001MNRAS.328..726S}, which we suitably changed to allow
extraction of bound structures from N-Body/SPH simulations also in the
presence of gas and star particles.  In Section~\ref{sec:sub}, we present our
results about the properties of the substructures in both DM and
hydrodynamical simulations. We summarize our results and outline our main
conclusion in Section \ref{sec:conc}. An Appendix is devoted to the
presentation of tests of resolution and numerical stability of our analysis.


\section{Simulations} \label{sec:sim}

In this section we describe our set of simulated clusters, the
different physical processes we considered, and the numerical choices
made for the different runs.

\begin{table*}
\begin{tabular}{|c|p{4cm}|p{3cm}|p{7cm}|}
\hline
Name & Included Physics & Reference & Used in \\
\hline
\hline
\cline{1-2}
\multicolumn{4}{l}{dark matter only} \\
\cline{1-2}
\hline

{\it dm} & gravity & -- & \citet{2005AA...442..405P,2008MNRAS.386.2135G,2007AA...461...25M,2008AA...482..403M}  \\
\multicolumn{4}{l}{
({\it g676}, {\it g914},{\it g1542}, {\it g3344},{\it g6212},
{\it g1}, {\it g8},{\it g51}, {\it g72} and
{\it g696})
}\\
\hline
\cline{1-2}
\multicolumn{4}{l}{non radiative runs}\\
\cline{1-2}
\hline

{\it ovisc} & usual parametrisation of artificial viscosity &
\citet{2005MNRAS.364..753D} &
\citet{2005AA...442..405P,2006MNRAS.365.1021E,2006MNRAS.370..656D,2007MNRAS.378.1248B,2008MNRAS.386...96C,
       2006AA...455..791P,2007AA...474..745P,2007AA...461...25M}\\
\multicolumn{4}{l}{
({\it g676}, {\it g914},{\it g1542}, {\it g3344},{\it g6212},
{\it g1}, {\it g8},{\it g51}, {\it g72} and
{\it g696})
}\\
\hline

{\it svisc} & artificial viscosity based on signal-velocity &
\citet{2005MNRAS.364..753D} & \\
\multicolumn{4}{l}{
({\it g676}, {\it g914},{\it g1542}, {\it g3344},{\it g6212},
{\it g1}, {\it g8},{\it g51} and {\it g72})
}\\
\hline

{\it lvisc} & time varying, low artificial viscosity scheme &
\citet{2005MNRAS.364..753D} & \citet{2006MNRAS.365.1021E,2006MNRAS.369L..14V} \\
\multicolumn{4}{l}{
({\it g676}, {\it g914},{\it g1542}, {\it g3344},{\it g6212},
{\it g1}, {\it g8},{\it g51} and {\it g72})
}\\
\hline

\cline{1-2}
\multicolumn{4}{l}{cooling, star-formation and thermal feedback} \\
\cline{1-2}
\hline

{\it csfnw} & no winds &
 -- & \citet{2006MNRAS.365.1021E,2006MNRAS.367.1641B}\\
\multicolumn{4}{l}{
({\it g676} and {\it g51})
}\\
\hline

{\it csf} & weak winds ($v_w=340$km/h) &
 -- &
\citet{2005AA...442..405P,2005ApJ...618L...1R,2006MNRAS.365.1021E,2007MNRAS.378.1248B,
2006MNRAS.367.1641B,2006MNRAS.369.1459A,2007MNRAS.382..397A}\\
\multicolumn{4}{l}{
({\it g676}, {\it g914},{\it g1542}, {\it g3344},{\it g6212},
{\it g1}, {\it g8},{\it g51}, {\it g72} and
{\it g696})
}\\
\hline

{\it csfsw} &  strong winds ($v_w=480$km/h) &
 -- & \citet{2006MNRAS.365.1021E,2006MNRAS.367.1641B}\\
\multicolumn{4}{l}{
({\it g676} and {\it g52})
}\\
\hline

{\it csfc} & weak winds ($v_w=340$km/h) and thermal conduction ($\kappa=0.3$) &
\citet{2004ApJ...606L..97D} & \citet{2006MNRAS.365.1021E,2006MNRAS.369.2013R,2007MNRAS.378.1248B}\\
\multicolumn{4}{l}{
({\it g676}, {\it g914},{\it g1542}, {\it g3344},{\it g6212},
{\it g1}, {\it g8},{\it g51}, {\it g72} and
{\it g696})
}\\
\hline
\end{tabular}
\caption{Symbolic names of the different runs analysed here, together
  with a short description of the physical processes included, and
  references to studies that used the simulations previously or that
  give specific details for the physical models used.  For each
  physical model, we also specify for which cluster the simulations
  have been carried out.}
\label{tab:physics}
\end{table*}

\begin{table*}
\begin{tabular}{|c|c|c|c|c|c|c|c|c|c|c|c|c|}
\hline
 & \multicolumn{6}{c|}{{\it dm}}
 & \multicolumn{6}{c|}{{\it ovisc}} \\

Cluster
 &  $R_{\rm vir}$ & $M_{\rm vir}$ & $R_{200}$ & $M_{200}$ & $R_{500}$ & $M_{500}$
 &  $R_{\rm vir}$ & $M_{\rm vir}$ & $R_{200}$ & $M_{200}$ & $R_{500}$ & $M_{500}$ \\


\hline

g676.a
   &   1.43   &   1.60   &   1.07  &   1.39  &   0.71   &   1.06
   &   1.40   &   1.53   &   1.06  &   1.33  &   0.71   &   1.03\\
\hline

g914.a
  &    1.50   &   1.84   &   1.09   &   1.44  &   0.71   &   1.06
  &    1.46   &   1.73   &   1.09   &   1.43  &   0.71   &   1.06\\
\hline

g1542.a
  &    1.40   &   1.53   &   1.043   &   1.29  &   0.69  &   0.93
  &    1.40   &   1.53   &   1.043   &   1.30  &   0.69  &   0.94\\
\hline

g3344.a
   &   1.44   &   1.64   &   1.07   &   1.40  &   0.73   &   1.10
   &   1.43   &   1.63   &   1.07   &   1.39  &   0.73   &   1.07\\
\hline

g6212.a
   &   1.43   &   1.61   &   1.06  &    1.33   &  0.70   &   1.00
   &   1.43   &   1.61   &   1.06  &    1.31   &  0.70   &   1.00\\
\hline
g51.a
   &   3.26   &   19.21   &   2.39   &   15.33   &   1.56  &   10.66
   &   3.26   &   19.03   &   2.37   &   15.30   &   1.57  &   11.16\\

\hline
g72.a
   &   3.29   &   19.63   &   2.40   &   15.59   &   1.57   &   10.93
   &   3.29   &   19.63   &   2.37   &   15.29   &   1.59   &   11.20\\

g72.b
  &    1.71   &   2.81   &   1.21   &   2.03   &  0.71   &   1.04
  &    1.71   &   2.79   &   1.26   &   2.24   &  0.74   &   1.19\\

\hline
g1.a
   &   3.40   &   21.86   &   2.54   &   18.86   &   1.73  &    14.61
   &   3.37   &   21.16   &   2.50   &   17.80   &   1.69  &    13.59\\

g1.b
  &    2.31   &   6.83  &    1.64  &    5.07   &   1.04  &    3.26
  &    2.26   &   6.43  &    1.67  &    5.36   &   1.06  &    3.43\\

g1.c
   &   1.67  &    2.61  &    1.23  &    2.09  &   0.79   &   1.40
   &   1.69  &    2.66  &    1.23  &    2.09  &   0.79   &   1.41\\

g1.d
  &    1.60  &    2.29  &   1.10   &   1.53  &   0.63  &   0.71
  &    1.57  &    2.14  &   1.07   &   1.40  &   0.61  &   0.66\\

g1.e
  &    1.27   &   1.14  &   0.94  &   0.94  &   0.59  &   0.56
  &    1.26   &   1.10  &   0.93  &   0.93  &   0.63  &   0.70\\

g1.f
  &    1.14  &   0.81  &   0.77  &   0.53  &   0.49  &   0.34
  &    1.11  &   0.77  &   0.77  &   0.53  &   0.50  &   0.34\\

\hline
g8.a
   &   3.90   &   32.70   &   2.86  &    26.63   &   1.90   &   19.64
   &   3.94   &   33.87   &   2.89  &    27.56   &   1.93   &   20.60\\

g8.b
  &    1.50   &   1.86   &   1.09   &   1.44   &  0.69  &   0.93
  &    1.54   &   2.01   &   1.14   &   1.70   &  0.74  &   1.20\\

g8.c
   &   1.36   &   1.37  &   0.93  &   0.93  &   0.60  &   0.60
   &   1.43   &   1.61  &   1.03  &   1.24  &   0.61  &   0.66\\

g8.d
   &   1.34   &   1.34  &   0.89  &   0.80  &   0.59  &   0.56
   &   1.40   &   1.51  &   1.00  &   1.14  &   0.66  &   0.81\\

g8.e
  &    1.30  &    1.23  &   0.96  &    1.00  &   0.60  &   0.61
  &    1.33  &    1.30  &   0.96  &    1.00  &   0.63  &   0.69\\

g8.f
  &    1.14  &   0.83  &   0.83  &   0.66  &   0.46  &   0.29
  &    1.20  &   0.94  &   0.87  &   0.76  &   0.54  &   0.47\\

g8.g
   &   1.11  &   0.76  &   0.77  &   0.53  &   0.47  &   0.31
   &   1.06  &   0.66  &   0.76  &   0.49  &   0.50  &   0.36\\

\hline
g696.a
  &    3.26   &   19.07   &   2.40   &   15.79   &   1.56   &   10.81
  &    3.24   &   18.94   &   2.39   &   15.51   &   1.57   &   11.00\\

g696.b
   &   3.13   &   17.04   &   2.23   &   12.64   &   1.43   &   8.23
   &   3.19   &   17.87   &   2.29   &   13.51   &   1.47   &   9.10\\

g696.c
   &   2.79   &   11.97   &   2.07   &   10.04  &    1.27  &    5.81
   &   2.77   &   11.87   &   2.07   &   10.13  &    1.29  &    6.06\\

g696.d
  &    2.67  &   10.57  &    1.83  &   7.04   &   1.14   &   4.30
  &    2.59  &    9.51  &    1.81  &   6.80   &   1.14   &   4.29\\

\hline
\end{tabular}
\caption{General properties of the  simulated
  clusters at $z=0$ for the cases of DM-only runs (left part of the
  table) and for the {\it ovisc} version of the non--radiative runs
  (right part of the table).  Column 1: cluster names; cols. 2 (7) and
  3 (8): virial radii (in units of Mpc) and virial
  masses (in units of $10^{14}M_\odot$); cols. 5 (10) and 4 (9): masses contained within
  the radius $R_{200}$, encompassing an average density of
  $200\,\rho_{\rm crit}$ and values of $R_{200}$; cols. 7 (12) and
  6 (11): the same as before, but referring to a mean density of
  $500\,\rho_{\rm crit}$.}
\label{tab:cluster}
\end{table*}

\begin{figure*}
\includegraphics[width=0.85\textwidth]{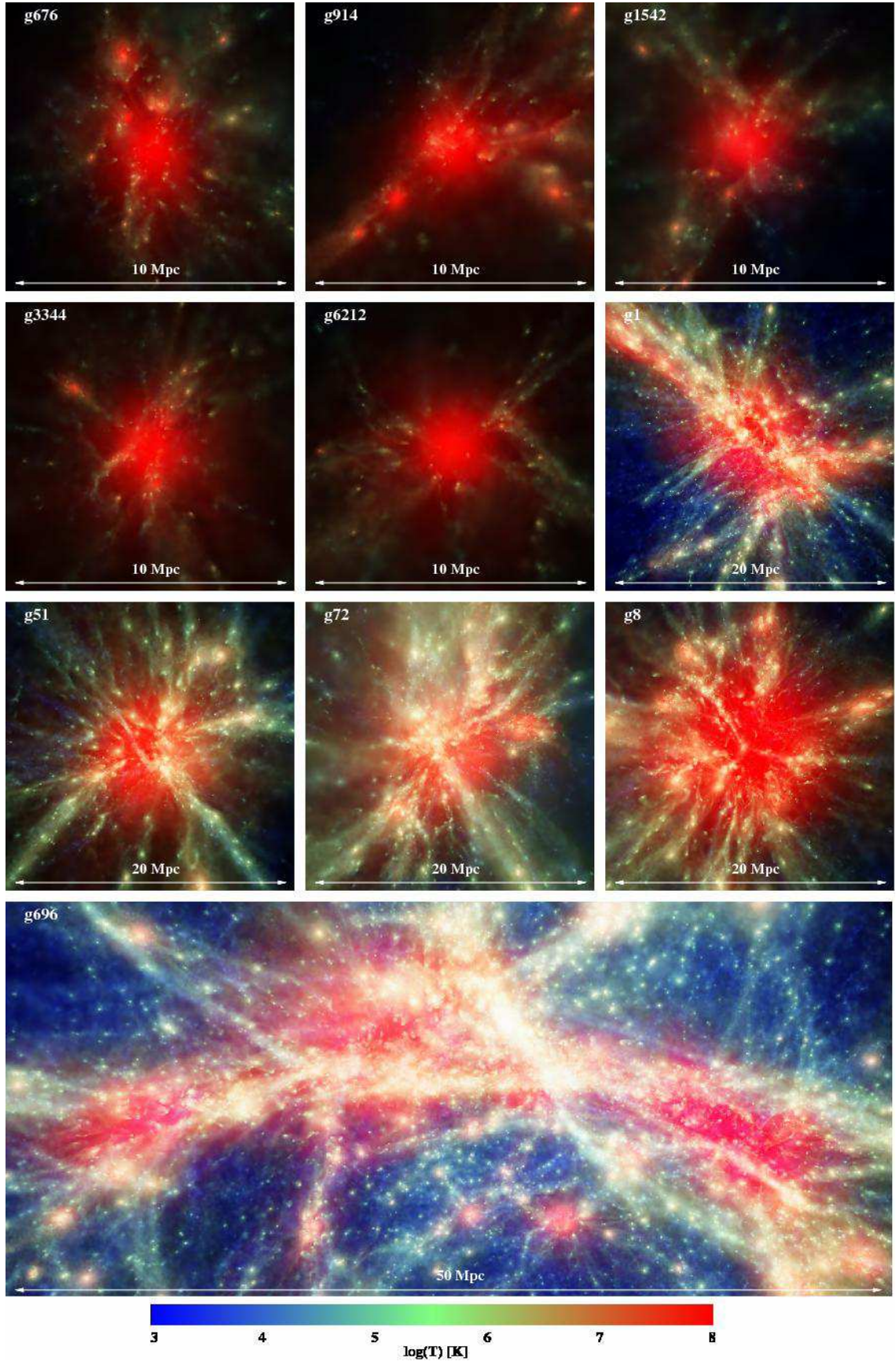}
\caption{Visualisation of the gas temperature in the
non--radiative ({\it ovisc}) simulations of the ten clusters
discussed in the text, carried out with the ray--tracing software
{\small SPLOTCH}. An animation flying through one of the simulated
clusters can be downloaded from {\tt
http://www.mpa-garching.mpg.de/galform/data\_vis/index.shtml\#movie9}.
} \label{fig:galery_t}
\end{figure*}

\subsection{Simulated physics} \label{sec:sim_phys}

The simulations were carried out with the TreePM/SPH code {\small
  GADGET-2} \citep{springel2001,springel2005}, which makes use of the
entropy--conserving formulation of SPH \citep{springel02}.  Some of
our non--radiative simulations were carried out with different
implementations of artificial viscosity compared with the standard one
in {\small GADGET-2}, however, which allows us to investigate the
effect of numerical viscosity on the stripping of gas from
substructures within galaxy clusters. In \citet{2005MNRAS.364..753D}
it was shown that the amount of turbulence detectable within the
cluster atmosphere is quite sensitive to the numerical treatment of
the artificial viscosity, so one expects that the stripping of the gas
from substructures might also be affected by these numerical details.
In the following, we will label simulations that use the original
parametrisation of the artificial viscosity by
\citet{1983JCoPh..52....374S,1995JCoPh.121..357B} as {\it ovisc}.
This was found to be the scheme with the highest numerical viscosity
in the study of \citet{2005MNRAS.364..753D}.  An alternative
formulation with slightly less numerical viscosity is based on the
signal velocity approach of \citet{1997JCoPh..136....298S}, and is
labelled as {\it svisc}. Finally, we considered a modified artificial
viscosity scheme as originally suggested by
\citet{1997JCoPh..136....41S}, labelled as {\it lvisc}. In this
scheme, every particle evolves its own time-dependent viscosity
parameter, with the goal to only have high viscosity in regions where
it is really needed.  In this scheme, shocks are generally as well captured
as in the standard approach, but regions away from the shocks
experience less artificial viscosity, such that an inviscid ideal gas
is represented more faithfully.  As a result, turbulence driven by
fluid instabilities can be better resolved, and simulated galaxy
clusters are able to build up a higher level of turbulence generated
along the shear flows arising in the cosmological structure formation
process \citep[see][]{2005MNRAS.364..753D}.

{\small GADGET-2} also includes, if enabled, radiative cooling,
heating by a uniform redshift--dependent UV background
\citep{1996ApJ...461...20H}, and a treatment of star formation and
feedback processes.  Simulations that account for radiative cooling
and star formation allow us to study the effect of a compact stellar
core at the centre of substructures on subhalo survival and
disruption.  The prescription of star formation we use is based on a
sub--resolution model to account for the multi--phase structure of the
interstellar medium (ISM), where the cold phase of the ISM is the
reservoir of star formation \citep{springel2003}. Supernovae heat the
hot phase of the ISM and provide energy for evaporating some of the
cold clouds, thereby leading to self-regulation of the star formation
and an effective equation of state for describing its dynamics.

As a phenomenological extension of this feedback scheme,
\citet{springel2003} also included a simple model for galactic winds,
whose velocity, $v_w$, scales with the fraction $\eta$ of the SN
type-II feedback energy that contributes to the winds. The total
energy provided by SN type-II is computed by assuming that they are
due to exploding massive stars with mass $>8\,{\rm M}_\odot$ from a
\cite{1955ApJ...121..161S} initial mass function (IMF), with each SN
releasing $10^{51}$ ergs of energy.  In our simulations with winds, we
will assume $\eta=0.5$ and 1, yielding $v_w\simeq 340$ ({\it csf}) and
480 km s$^{-1}$ ({\it csfsw}), respectively, while we will also
explore the effect of switching off galactic winds ({\it csfnw}). We
refer to Table~\ref{tab:physics} for a schematic description and
overview of the physical and numerical schemes included in our
simulations.

\begin{figure*}
\includegraphics[width=0.9\textwidth]{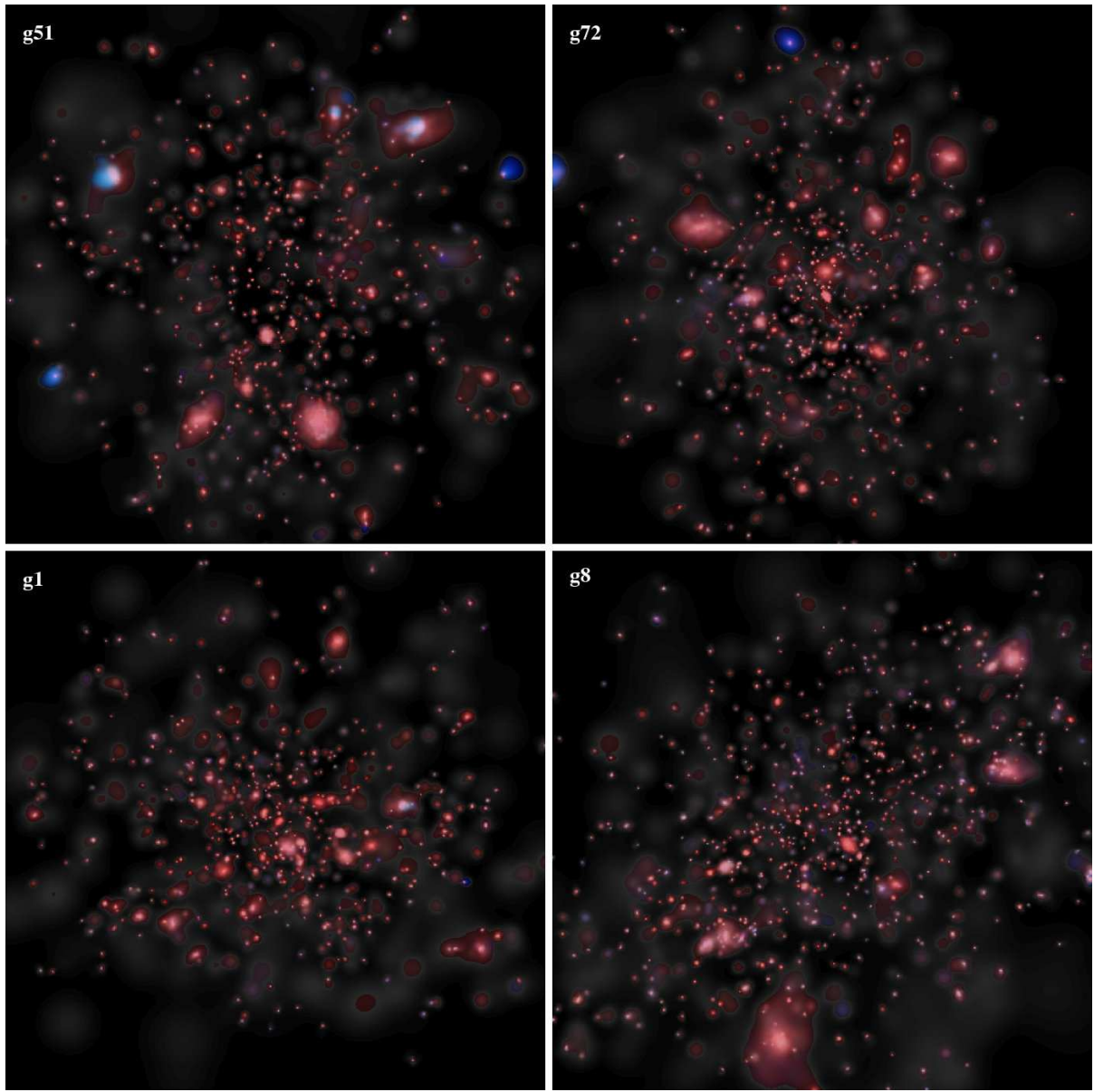}
\caption{Visualisation of all the identified galaxies within the
  virial radius of the g51, g72, g1 and g8 clusters (from top left to
  bottom right), simulated with cooling and star-formation and
  galactic winds ({\it csf}). About 1000 galaxies are identified
  within the virial radius of each of these massive clusters. Only the
  self-bound gas and star particles within substructures are included in the
  visualization, therefore excluding the hot atmosphere of the cluster and
  both the diffuse intracluster
  stellar population and the stars associated with the BCG. The
  colours of the stars represent their age, using a red to light blue
  colour-table for decreasing age. Only a few of these galaxies, about
  5-10 per cluster, still carries a self--bound hot gas halo, and
  those are located in the cluster peripheries.  In the visualisation
  they appears as dark blue, extended halos, often with a comet--like
  shape, stretched away from the galaxies.} \label{fig:galaxies}
\end{figure*}

For some of our cluster simulations we also included the effect of
heat conduction ({\it csfc}), based on the implementation described by
\citet{2004MNRAS.351..423J}. In the simulations presented here, we
assume an isotropic effective conductivity parameterized as a fixed
fraction of $1/3$ the Spitzer rate. We also account for saturation,
which can become relevant in low--density gas and at the interface
between cold and hot media. We refer to \citet{2004ApJ...606L..97D}
for more details on the effect of thermal conduction on the
thermodynamical properties of the intra--cluster medium (ICM).

\subsection{The set of simulated clusters}

The clusters analyzed in this study are extracted from 10
re-simulations of Lagrangian regions selected from a cosmological
lower resolution DM-only simulation \citep{2001MNRAS.328..669Y}. This
parent simulation has a box--size of $479\hm$, and assumed a flat
$\Lambda$CDM cosmology with $\Omega_m=0.3$ for the matter density
parameter, $H_0=70 \vel$Mpc$^{-1}$ for the Hubble constant, $f_{\rm
  bar}=0.13$ for the baryon fraction and $\sigma_8=0.9$ for the
normalisation of the power spectrum. As such, the values of $\Omega_m$
and $\sigma_8$ in this model are somewhat higher than the
best--fitting values obtained from the analysis of the 5-year WMAP
data \citep{2008arXiv0803.0586D}, but are still consistent with the
current cosmological constraints.

Five of the regions have been selected to surround low--mass clusters
($m_{\rm cluster} \approx 10^{14}M_\odot$), while four other regions
surround massive ($m_{\rm cluster} \magcir 10^{15}M_\odot$)
clusters. Besides the central massive cluster, three of these four
regions also contain 12 additional smaller clusters, all having virial
masses above $10^{14}M_\odot$.  The tenth simulated region surrounds a
filamentary structure which includes four massive ($m_{\rm cluster} >
10^{15} M_\odot$) clusters \citep[see
][]{2006MNRAS.370..656D}. Ray--tracing images, obtained with the
{\small SPLOTCH} package \citep{Dolag08}, of the gas temperature for
the simulated cluster regions are shown in
Figure~\ref{fig:galery_t}. Besides giving an impression of the
complexity and the dynamics of the ICM within the simulated galaxy
clusters, this figure also highlights a couple of interesting key
features of hydrodynamical cluster simulations. The white colours in
general correspond to high density, whereas the red colour marks the
hot, shock-heated atmosphere of the cluster. The transition to the
un-shocked material (mainly indicated by the blue colour) is clearly
visible and indicates the location of the accretion shocks, which for
clusters at redshift zero are typically at much larger distances than
the virial radius (see Table~\ref{tab:cluster}). Also clearly visible
are denser and colder filaments that penetrate the hot cluster
atmosphere. Many of the gaseous halos of the substructures show comet
like features, a tell tale sign of gas being stripped. Sometimes this
happens already at relatively large distances from the cluster center,
demonstrating a change in the enviroment that becomes important
already at distances of several virial radii
\citep[e.g.][]{2006MNRAS.370..656D}.

Using the ``Zoomed Initial Conditions'' (ZIC) technique
\citep{tormen97}, these regions were re-simulated with higher mass and
force resolution by populating their Lagrangian volumes with a larger
number of particles, while appropriately adding additional
high--frequency modes drawn from the same power spectrum. To optimise
the setup of the initial conditions, the high resolution region was
sampled with a $16^3$ grid (for the more massive systems even with a $64^3$
grid), where only sub-cells are re-sampled at high resolution to allow
for nearly arbitrary shapes of the high resolution region. The exact
shape of each final high--resolution region was iteratively obtained
by repeatedly running dark-matter only simulations until the target
objects were free of any contamination by lower--resolution boundary
particles out to 3-5 virial radii. The initial unperturbed particle
distribution (before imprinting the Zeldovich displacements) was
realized through a relaxed glass-like configuration
\citep{1996clss.conf..349W}.

Gas was then added to the high-resolution regions by splitting
each parent particle into a gas and a DM particle. The gas and the
DM particles were displaced by half the original mean
inter-particle distance, such that the centre-of-mass and the
momentum of the original particle are conserved. As a further
optimisation to save CPU time, the splitting of the DM particles
in our largest simulations (i.e., those of the filamentary
structure) was applied only to a subset of the high-resolution
particles, selected by the following procedure. Instead of
defining the high resolution region with the help of a grid, we
used a surface around the centre of the high resolution region.
For this purpose, we pixelised a fiducial sphere around the centre
of the high resolution region using $4\times 16^2$ pixels, making
use of the {\small HEALPIX} tessellation of the full sky \citep{healpix},
which has the advantage of offering an equal solid angle covered
by each pixel. We then traced back in Lagrangian space all the
particles in the DM-only run which at $z=0$ end up within 5 virial
radii of the four massive clusters. All such particles were then
projected onto the spherical coordinate system, where we recorded
the maximum distance within each pixel of our {\small HEALPIX} representation.
In a second step, all particles within the high resolution region
were projected onto the spherical coordinate system and only those
with distances smaller than the recorded maximum distance of the
corresponding pixel of our {\small HEALPIX} representation were then marked to
define the region where to add gas. This optimisation procedure
reduced the number of gas particles by about 40 per cent, without
significantly reducing the size of the regions around the target
objects where hydrodynamics is correctly computed.

The final mass resolution of the gas particles in our simulations
is $m_{\rm gas}=2.4\times 10^8\,{\rm M}_\odot$. Thus, the massive
clusters were resolved with between $2\times10^6$ and
$4\times10^6$ particles. In all simulations, the gravitational
softening length was kept fixed at $\epsilon=42\,\mathrm{kpc}$
comoving Plummer-equivalent, and was switched to a physical
softening length of $\epsilon=7\,h^{-1}\mathrm{kpc}$ at $z=5$.

The high resolution regions around the massive clusters are quite
large and therefore also include other lower mass systems, which are
still free of contamination from low--resolution particles.  Hence,
these systems can be included in our analysis. In this way, we end up
with a total sample of 25 clusters having a mass of at least
$10^{14}{\rm M}_\odot$.  The basic characteristics of these cluster,
such as masses and radii at different overdensities, both for the
DM--only runs and the non--radiative {\it ovisc} runs, are given in
Table~\ref{tab:cluster}.

To study resolution effects in the substructure mass distribution in
the DM runs, we also performed all these simulations at six times
higher mass resolution, with the gravitational softenings decreased
accordingly by a factor $6^{1/3}\simeq 1.8$. In the following, we
refer to these simulations as our high-resolution DM runs.

Finally, In Appendix A we will show results of a resolution study of
one single cluster, taken from a different set of cluster simulations
\citep{2006MNRAS.367.1641B}, which has been simulated at three
different resolution. In the same Appendix we will also discuss the
effect of adding gas particles in the initial conditions in such a way
that there are 8 times fewer gas particles, but each having roughly
the mass as that of the high-resolution DM particles.


\section{Detection of substructures} \label{sec:detection}

Substructures within halos are usually defined as locally
over-dense, self-bound particle groups identified within a larger
parent halo. In our analysis, the identification of these
substructures is performed by applying the {\small SUBFIND}
algorithm \citep{2001MNRAS.328..726S}, which is defined in its
original form only for dark matter only simulations. In brief, as
a first step, we employ a standard friends-of-friends (FoF)
algorithm to identify the parent halos. In our analysis we have
used a FoF linking length of 0.16 times the mean DM particle
separation. Note that this linking length is obtained when scaling
the {\it standard} linking length of $0.2$ by
$~(\Delta_c/\Omega)^{-1/3}$ according to the adopted cosmology and
leads to groups having the overdensity characteristic of
virialized objects predicted by the spherical collapse model
\citep[see][]{1996MNRAS.282..263E}. The density of each particle
within a FoF group is then estimated by adaptive
kernel-interpolation, using the standard SPH approach with a
certain number of neighbouring particles. Using an excursion set
approach where a global density threshold is progressively
lowered, we find locally overdense regions within the resulting
density field, which form a set of substructure candidates. The
outer `edge' of the substructure candidate is determined by a
density contour that passes through a saddle point of the density
field; here the substructure candidate joins onto the background
structure.  In a final step, all substructure candidates are
subjected to a gravitational unbinding procedure where only the
self-bound part is retained. If the number of bound particles left
is larger than a prescribed minimum detection threshold, we
register the substructure as genuine {\it subhalo}.

For full details on this substructure--finding algorithm as applied to dark
matter only simulations we refer the reader to the paper by
\citet{2001MNRAS.328..726S}. In the following, we briefly describe the
modifications to the algorithm that we implemented in order to make it
applicable to simulations which also contain gas and star particles.

Because the dark matter particles and the baryonic particles are in general
distributed differently (especially in simulations with cooling), it is
problematic to apply the FoF algorithm to all particles at once on an equal
footing. This would not select groups that are bounded by a clearly defined
density contour, and produces systematic biases in the relative mass content
of baryons and dark matter. We therefore apply the FoF only to the dark matter
component of a simulation, using the same linking length that would be applied
in a corresponding DM-only run. This selects halos that are bounded at
approximately the same dark matter overdensity contour, independent of the
presence or absence of baryonic physics.  Each gas and star particle is then
associated with its nearest DM particle, i.e.~if this DM particle belongs to a
FoF group, then the corresponding baryonic particle is also associated with
that group. This effectively encloses all baryonic material as part of a FoF
group that is contained in the original dark matter overdensity contour.

We only keep FoF groups containing at least 32 DM particles for further
analysis. Because a significant fraction of such small dark matter haloes are
spurious, it is better to impose the detection limit just on the dark matter
particle number, and not on the total number or total FoF group mass, because
the latter would boost the number of spurious substructures detected in
hydrodynamical runs. Also, this limit ensures that the halo number densities
detected in DM-only and in hydrodynamic runs should only differ because of the
effects of the baryonic component on the dark matter dynamics, and not because
of a change in the group detection procedure.

The second adjustment we made in {\small SUBFIND} is the procedure for density
estimation in the presence of further components besides dark matter. We
address this by first estimating the densities contributed by DM, gas
particles and star particles at a given point separately for each of the
components and then adding them up, i.e.~the density carried by each of the
three species is computed with the SPH kernel interpolation technique for
neighbouring particles of the same species only.  Again, this is done to avoid
possible systematic biases in the density estimates. If they are done
with all particles at once, the different particle masses and spatial
distributions of the particle types can introduce comparatively large errors,
for example when a few heavy dark matter particles dominate the sum over
neighbours that mostly consist of light star particles. We note that we have
carried out all the density computations with 32 neighbours.

Typically, the density contribution from the gas component is slightly
smoother than that from the DM component (thanks to the gas pressure), whereas
star particles generally trace a highly concentrated density field, thus
making it easier to detect substructures dominated by star particles. As a net
result, we find that the resulting total density field is a bit more noisy
than the density field in pure DM runs.  This led us to slightly modify {\small
  SUBFIND}'s procedure to detect saddle points in the density field, and hence
substructure candidates. In the default version of the algorithm, this is done
by looking at the nearest two particles of a point that is added to the
density field. We changed this to the nearest three points, finding that this
leads to a significant improvement of the robustness of the
identification of substructures in dissipative simulations while not changing
the behaviour of the algorithm in pure DM or non--radiative runs.

Finally, as a further refinement in {\small SUBFIND}, we take into account the
internal thermal energy of the gas particles in the gravitational unbinding
procedure.  After unbinding, we keep substructures in our final list of
subhalos if they contain a minimum of 20 dark matter {\em and} star particles.
We ignore gas particles in this detection threshold in order to avoid
detection of spurious gravitationally bound gas clumps. In principle, a
cleaner mass threshold for detection would be obtained by also ignoring the
star particle count for this validation, but including them allows us to
identify as genuine self--bound substructures also very low mass systems that
are dominated by the stellar component.  Since stars form at the bottom of the
potential wells and are highly concentrated relative to the other two
components, we find this does not produce a significant component of spurious
subhalos.

For each of the FoF groups that correspond to a cluster, we obtain
in this way a catalogue of gravitationally bound subhalos that in
general contain a mixture of dark matter, star and gas particles.
The centre of each subhalo is identified with the minimum of the
gravitational potential occuring among the member particles.
Similarly, for the cluster as a whole, we determine a suitable
centre as the position of the particle that has the minimum
gravitational potential.  Around this point, we calculate the
virial radius and mass with the spherical-overdensity approach,
using the over-density predicted by the generalized spherical
top-hat collapse model \citep[e.g.][]{1996MNRAS.282..263E}. We
define as subhalos of a cluster all the substructures which are
identified within its spherical virial radius. Note that this can
sometimes include subhalos that belong to a FoF group different
from that of the main halo, due to the aspherical shapes of the
FoF groups. Likewise, not all of the subhalos in the cluster's FoF
group are necessarily inside its virial radius.

\begin{table}
\begin{tabular}{|l|c|c|}
\hline
Cluster Sample& $N_{-4}$ & $\alpha$ \\
\hline
low mass                    &  $141 \pm 106$ &  $-1.01 \pm 0.26$ \\
low mass (high resolution)  &  $131 \pm  32$ &  $-0.97 \pm 0.15$ \\
high mass                   &  $136 \pm  17$ &  $-1.00 \pm 0.07$ \\
high mass (high resolution) &  $142 \pm  22$ &  $-0.99 \pm 0.03$ \\
\hline
\end{tabular}
\caption{The best--fit parameters for the cumulative subhalo mass
  function of Eq.~(\ref{eqn:fit}). Reported here are the values for
  the low-- and high--mass samples of DM--only simulations. Results
  are shown for the simulations performed both at the standard and at
  the high resolution. Quoted errors correspond to the {\em rms}
  scatter computed among the best--fitting values obtained for the
  individual clusters.}
\label{tab:fit_dm}
\end{table}

\begin{figure*}
\begin{center}
\includegraphics[width=0.33\textwidth]{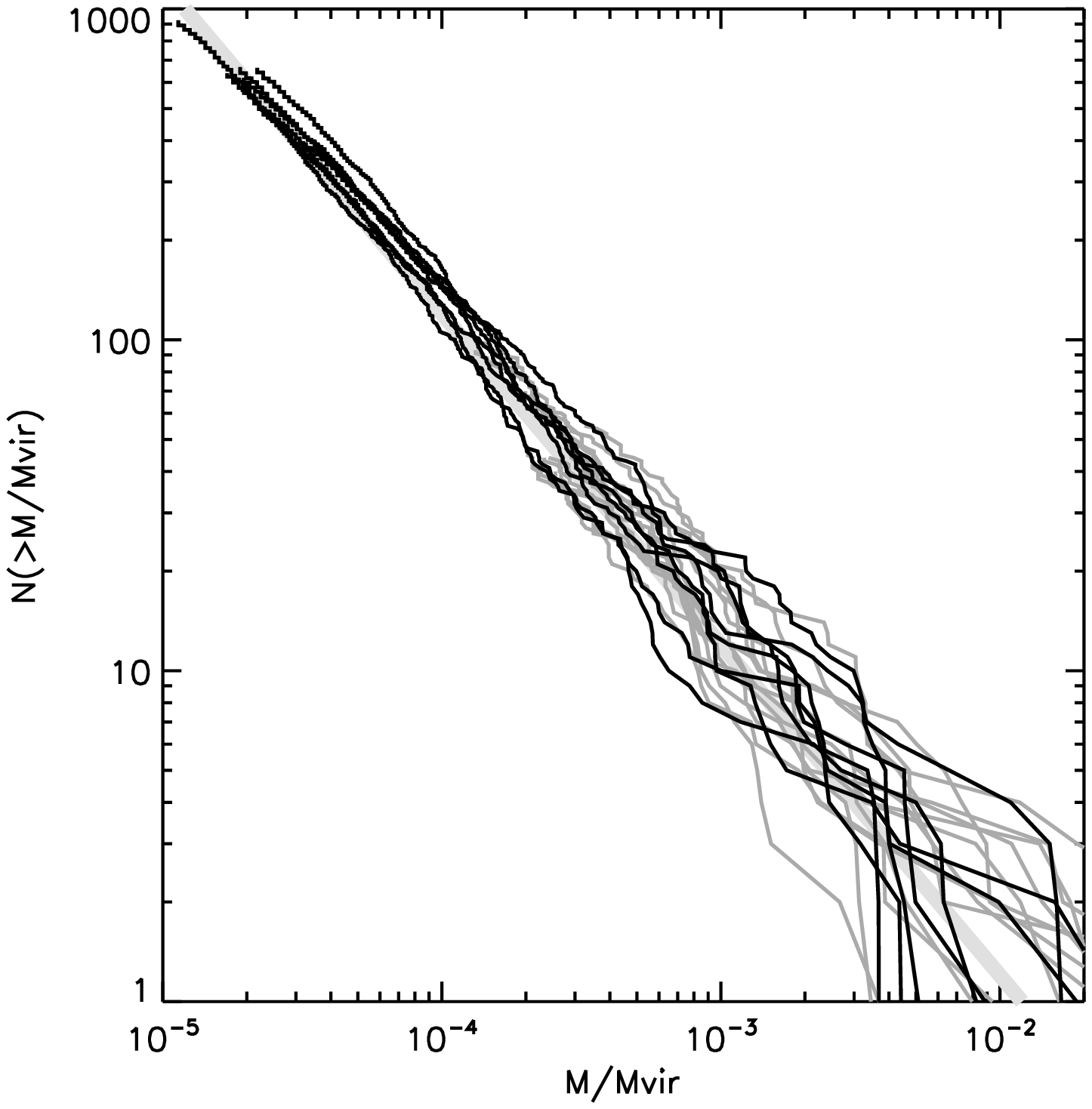}
\includegraphics[width=0.33\textwidth]{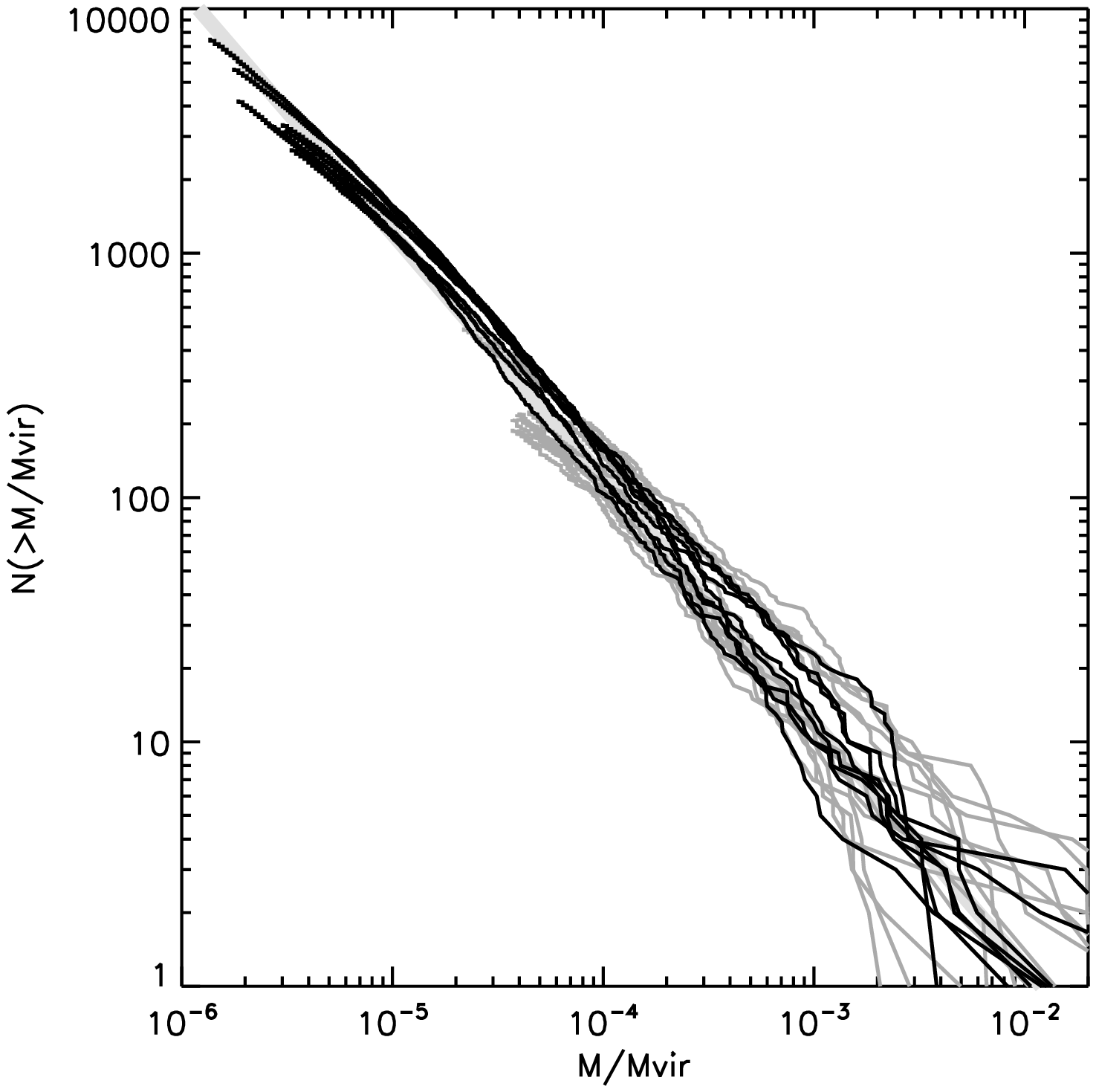}
\includegraphics[width=0.33\textwidth]{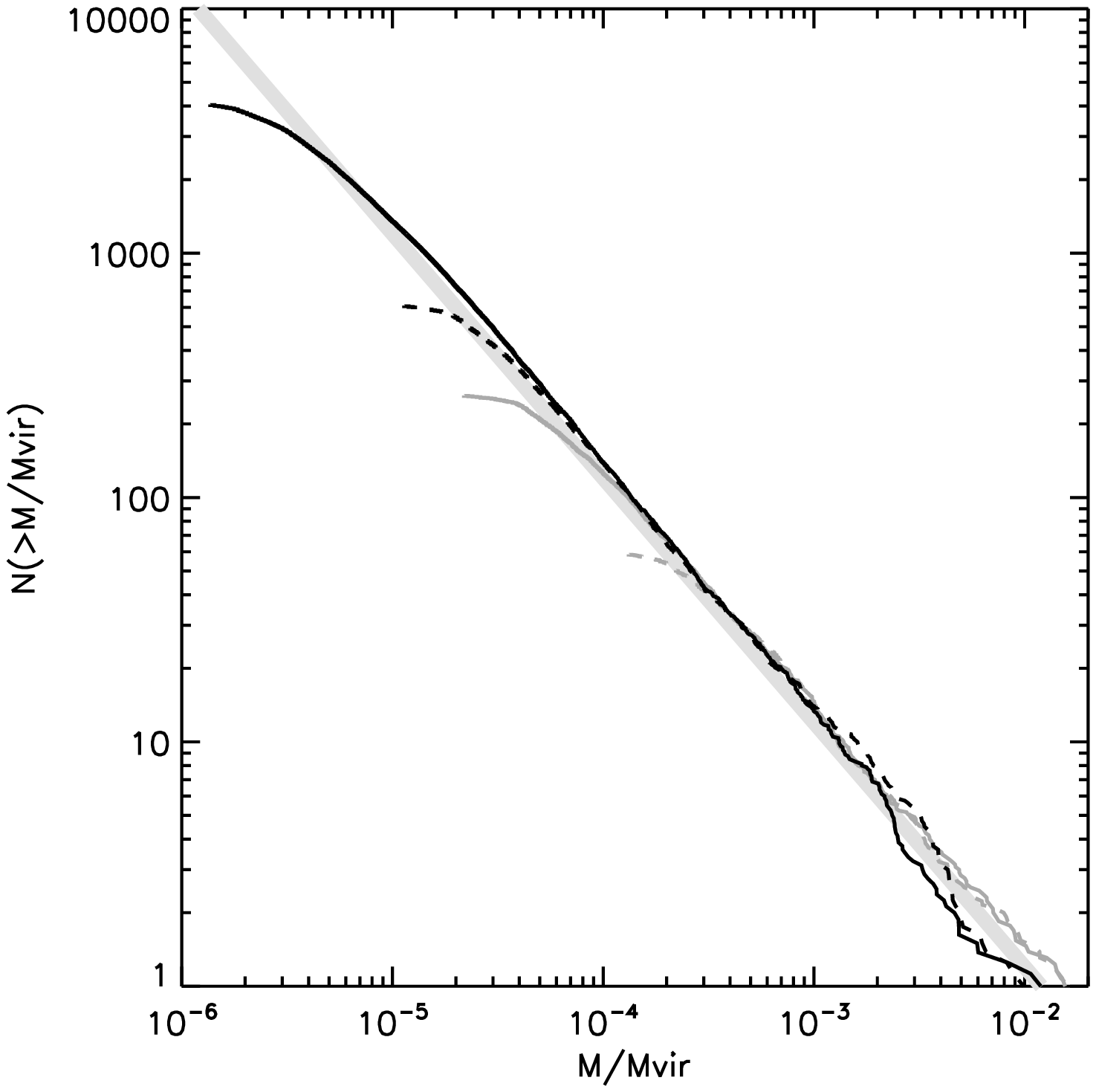}
\end{center}
\caption{The cumulative subhalo mass functions for the DM runs of our
  clusters, normalized to the virial mass of each cluster. Left panel:
  comparison between the 13 low mass halos (in grey) and the 8 high
  mass halos (in black).  Central panel: the same as in the left
  panel, but for the runs with six times higher mass resolution. Right
  panel: the average cumulative subhalo mass function for the low (in
  grey) and high (in black) mass sample from the DM-only runs. The
  dashed lines are for the simulations at the standard resolution
  while the solid lines are for the higher resolution runs. The
  corresponding best--fit values of the power--law fit to each of
  these curves are given in Table~\ref{tab:fit_dm}. For reference, the
  grey thick line marks a power-law with slope
  $-1$.}\label{fig:fig_mf_dm}
\end{figure*}

\section{Subhalos in galaxy clusters} \label{sec:sub}

\begin{figure*}
\includegraphics[width=0.33\textwidth]{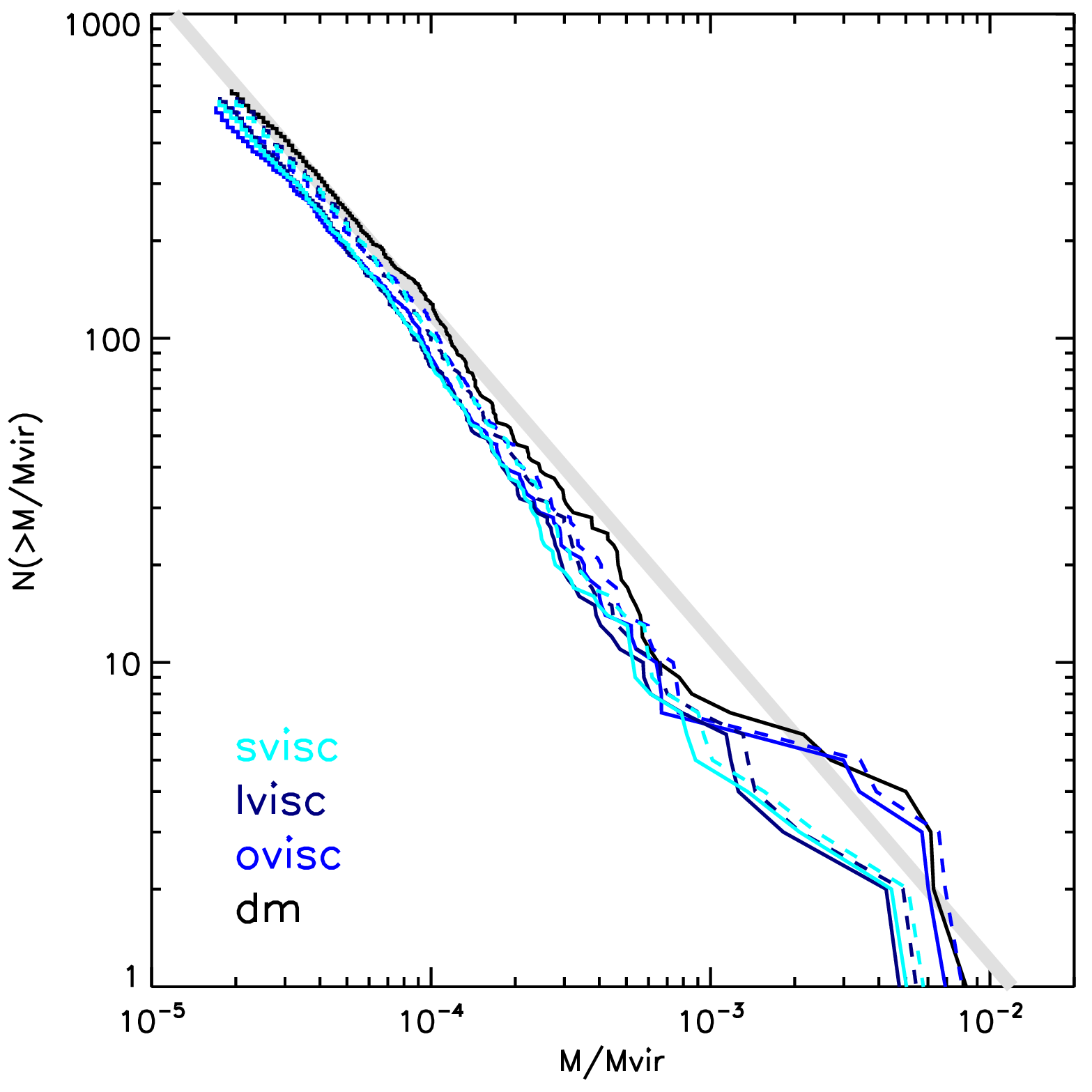}
\includegraphics[width=0.33\textwidth]{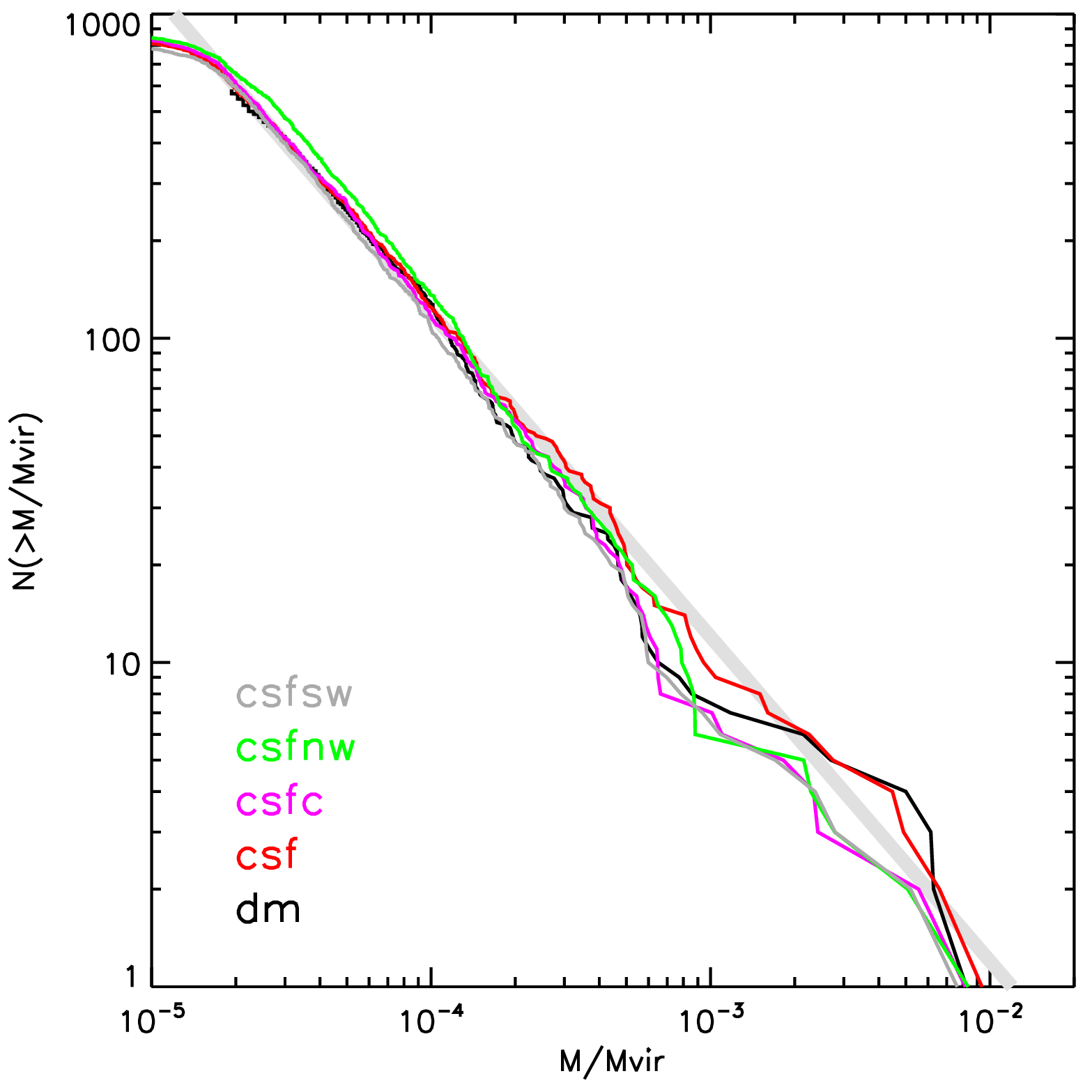}
\includegraphics[width=0.33\textwidth]{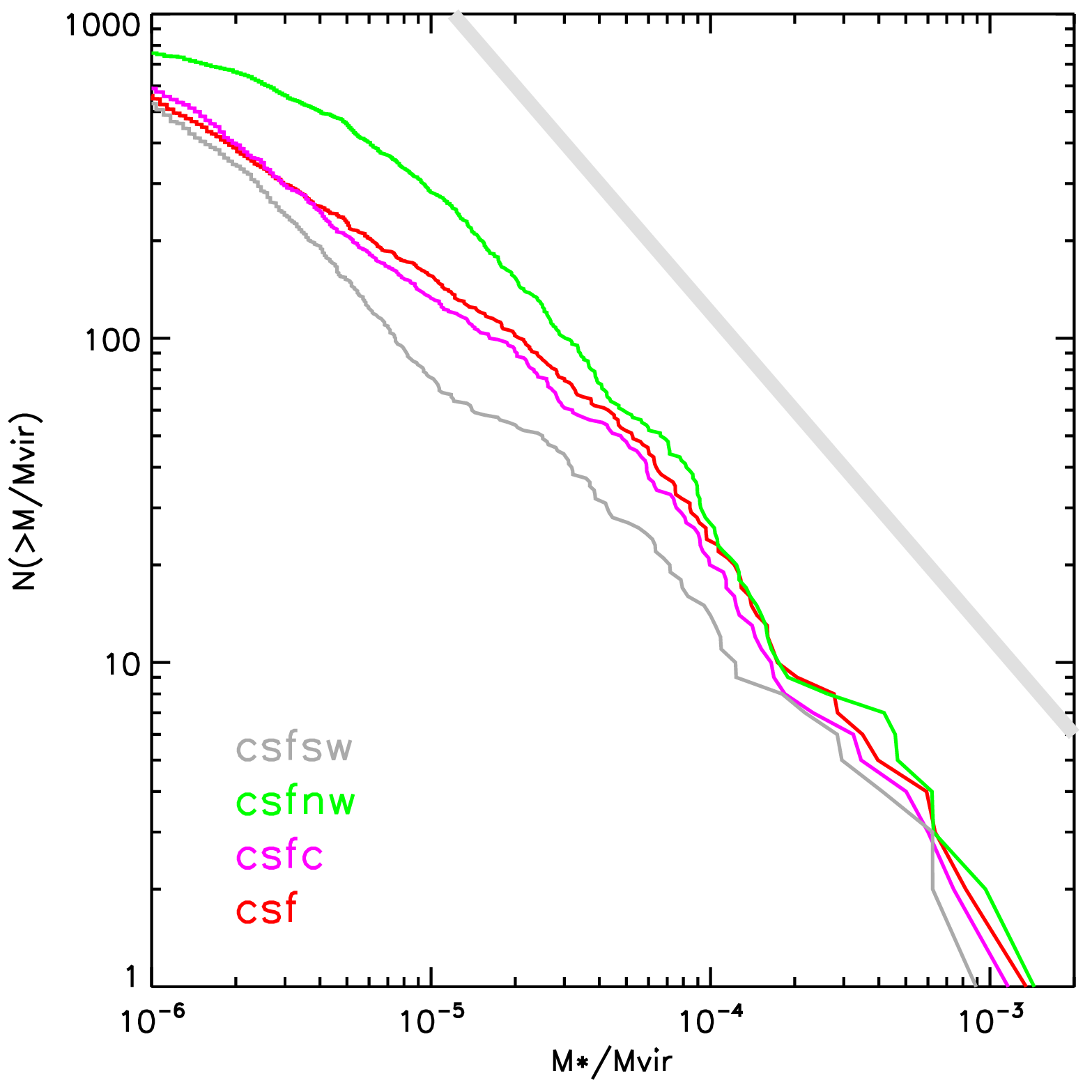}
\caption{The cumulative total mass functions (left and central panel)
  and stellar mass functions (right panel) for the subhalos identified
  in the $g51$ cluster at $z=0$, as obtained for different runs.  Left
  panel: comparison of the total subhalo mass functions of the DM run
  with the non--radiative runs corresponding to different schemes for
  the artificial viscosity (solid lines). 
  The dashed lines are obtained by correcting the mass of each
  sub--halo in the non--radiative runs by increasing it by the
  cosmological baryon fraction. 
  Central panel: the same as the left
  panel, but for the different radiative runs.  Right panel:
  comparison between the stellar mass functions obtained for the
  different radiative runs. As in Figure~\ref{fig:fig_mf_dm}, the
  thick grey line marks a power--law with slope $-1$.}
\label{fig:fig_phys}
\end{figure*}

\begin{figure*}
\includegraphics[width=0.33\textwidth]{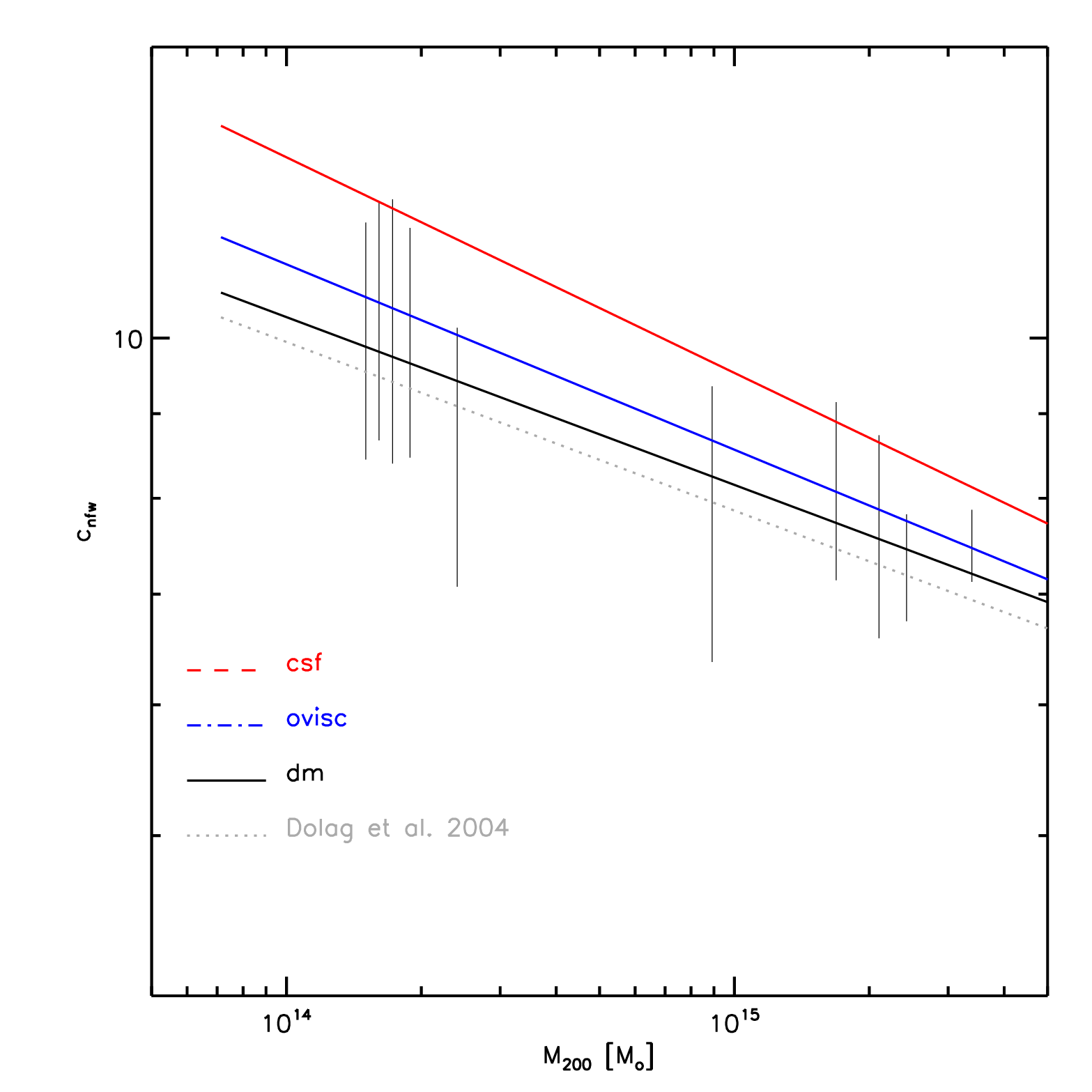}
\includegraphics[width=0.33\textwidth]{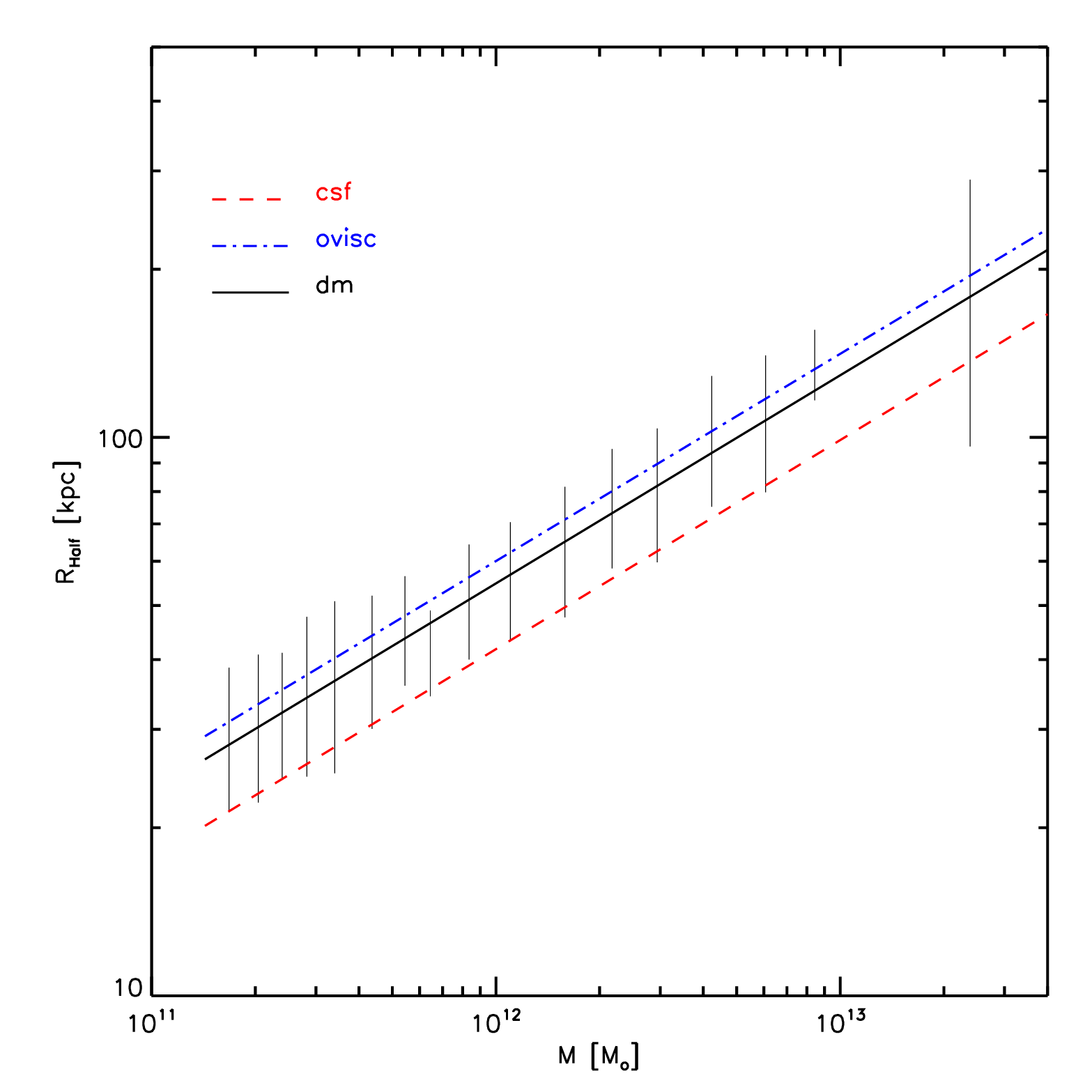}
\caption{Left panel: the concentration parameter, obtained by
    fitting a NFW profile \citep{1996ApJ...462..563N} to the dark
    matter density profiles in the hydrodynamical simulations, as a
    function of $M_{200}$. Here we used all halos with masses above
    $10^{14} M_\odot$ identified in all the simulation snapshots at
    redshift above $z=0.5$, thereby scaling the obtained concentration
    parameter by $(1+z)$ to compensate for the evolution. The
    different lines mark the results for the dark matter run ({\it
      dm}), the non-radiative run ({\it ovisc}) and the run with cooling
    and star formation ({\it csf}).  The dashed line gives the fit
    proposed in \citet{2004A&A...416..853D}. The right panel shows the
    half mass radius as a function of the sub-halo mass for all
    substructures in all the clusters at redshift $z=0$. The different
    lines distinguish the different simulations as before. In both
    panels the errorbars (shown only for the {\it dm} run)
    correspond to the {\em rms} scatter among all the clusters within
    each mass bin.}
\label{fig:fig_new}
\end{figure*}

\subsection{DM-only simulations} \label{sec:sub_dm}

As a first step, we characterize the mass distribution of subhalos for the DM
runs and check the effect of the mass of the parent halo and of
resolution. For this purpose, we split our sample of simulated clusters into a
high mass sample, containing 8 halos with virial masses above
$10^{15}{\rm M}_\odot$, and a low mass sample, containing 13 halos with masses
ranging between $\approx 1.0 \times 10^{14} {\rm M}_\odot$ and $\approx 3.0 \times
10^{14} {\rm M}_\odot$ (see Table~\ref{tab:physics}).

Figure \ref{fig:fig_mf_dm} shows the cumulative subhalo mass function,
normalized to the virial mass of the clusters (i.e. subhalo masses are
measured in units of the virial mass of the parent halo).  In the left panel
and in the central panels, the lines are for individual clusters, with grey
and black lines showing the low-mass and high-mass samples, respectively. The
scatter among different clusters is substantial for substructure counts below
$\siml 100$, which only includes the most massive structures. However, once a
few hundred substructures are considered, the abundance per unit virial mass
becomes quite uniform. There appears to be no significant difference in the
scatter of the high-mass end of the subhalo mass functions between our low-
and high-mass cluster samples.

In the middle panel, we show the same results as in the left panel, but now for
6 times higher mass resolution. This allows us to count down to subhalos that
are 6 times smaller in mass, yielding significantly higher total subhalo
counts. There is no clear difference in the cluster-to-cluster scatter
relative to the lower resolution simulations, suggesting that the scatter is
not due to numerical noise but rather reflects the genuine variations in the
abundance of massive subhalos in different halos.


This is corroborated by the right panel of Fig.~\ref{fig:fig_mf_dm}, where we
compare the average mass functions for the two subsets of low--mass and
high--mass clusters, and at the two numerical resolutions. The thick grey line
marks a power law with slope of $-1$, for comparison. Clearly, the average
slope of the subhalo mass function of both cluster samples and at both
resolutions agree perfectly.

This result is good in agreement with previous findings \citep[e.g.][]
{2004MNRAS.352L...1G,2004MNRAS.348..333D}. However, unlike
\citet{2004MNRAS.352L...1G}, we here do not find a significant trend of the
amplitude of the subhalo mass function as a function of halo mass.  a results
which is in line with the findings by \citet{2004MNRAS.348..333D}. However, it
may simply be the relatively small range of halo masses probed by our
simulations that prevents us from seeing the weak mass trend, despite the good
resolution reached in our most massive systems.  We note that in the
high--resolution versions of the high-mass clusters, each main halo is
resolved with more than 10 million dark matter particles within $R_{\rm vir}$,
allowing several thousand subhalos to be identified within each main halo.

A good quantitative fit to our measured  subhalo mass
functions can be obtained with a power-law of the form
\begin{equation}
   N_{m} = N_{-4} \left(\frac{m_{\rm sub}/M_{\rm vir}}
   {{\rm M}_\odot}\right)^\alpha .
\label{eqn:fit}
\end{equation}
We obtain values of the slope $\alpha$ and of the normalization $N_{-4}$ for
each cluster by performing a fit to the numerical data above a lower limit of
$100$ DM particles (to be insensitive to the resolution limit) and for $m_{\rm
  sub}/M_{\rm vir} < 0.01$ (in order to reduce sensitivity to the scatter in the
high-mass end of the subhalo mass function).  The resulting fitting parameters
are reported in Table~\ref{tab:fit_dm}, along with mean values over our
cluster sample and the errors due to the object-by-object scatter.  The
results of this table confirm that at high resolution we obtain rather stable
results for both the slope and the normalisation of the subhalo mass
functions.  Furthermore, we find no significant differences between the mass
functions of high and low mass clusters, both in slope and in normalization.

\subsection{Hydrodynamical runs} \label{sec:sub_gas}

We now turn to an analysis of the subhalo mass functions in our hydrodynamical
simulations.  As described in Section~\ref{sec:detection}, we have modified
the subhalo detection algorithm {\small SUBFIND} such that it can properly
take into account the presence of gas and star particles, besides the dark
matter particles. Our modifications have been guided by the desire to avoid
possible systematic biases due to the presence of multiple particle
components, such that the final subhalo catalogues can be directly compared
with those of a DM-only simulation, except that the subhalos can now also
contain gas and star particles.

\begin{figure*}
\includegraphics[width=0.33\textwidth]{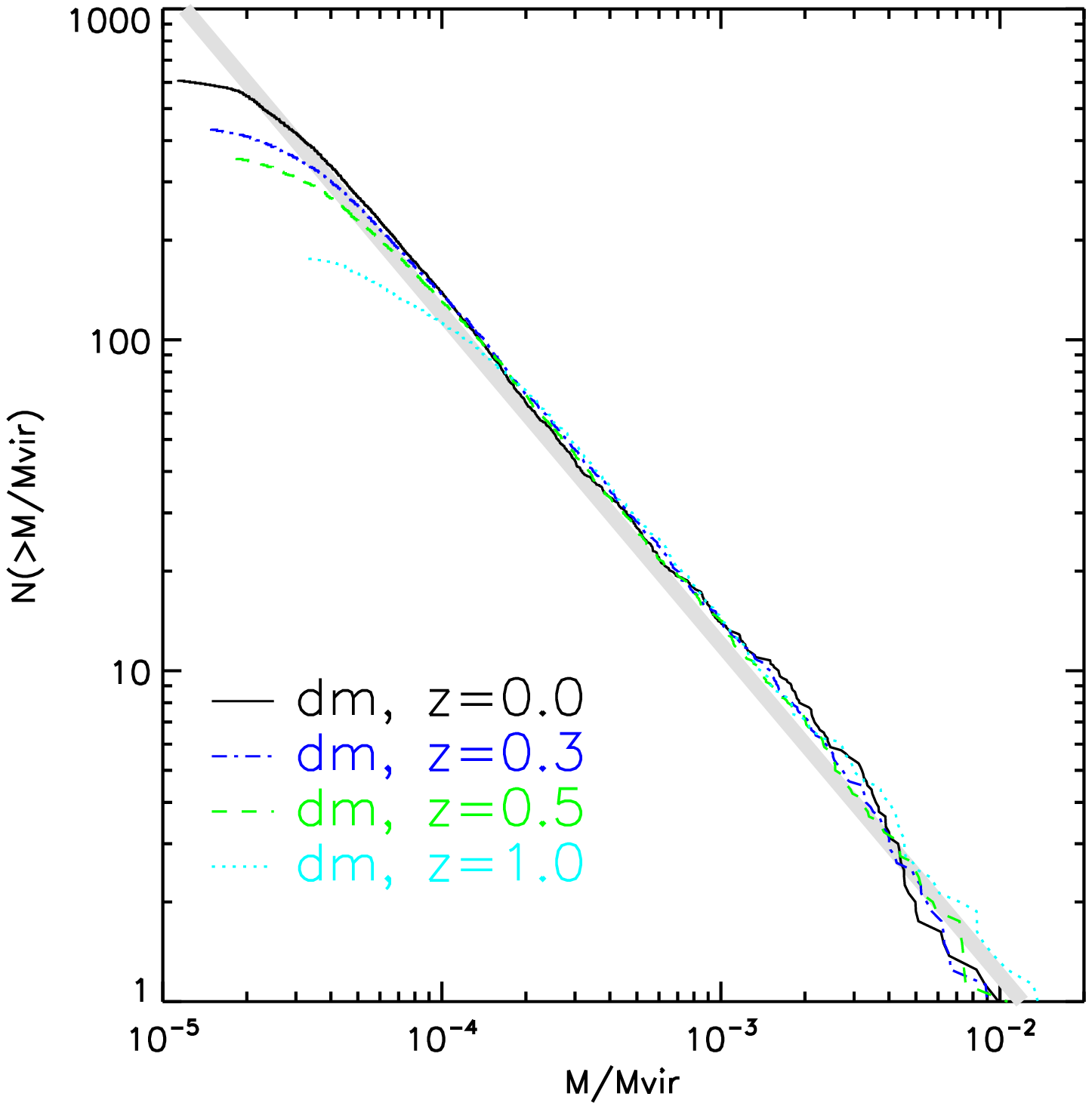}
\includegraphics[width=0.33\textwidth]{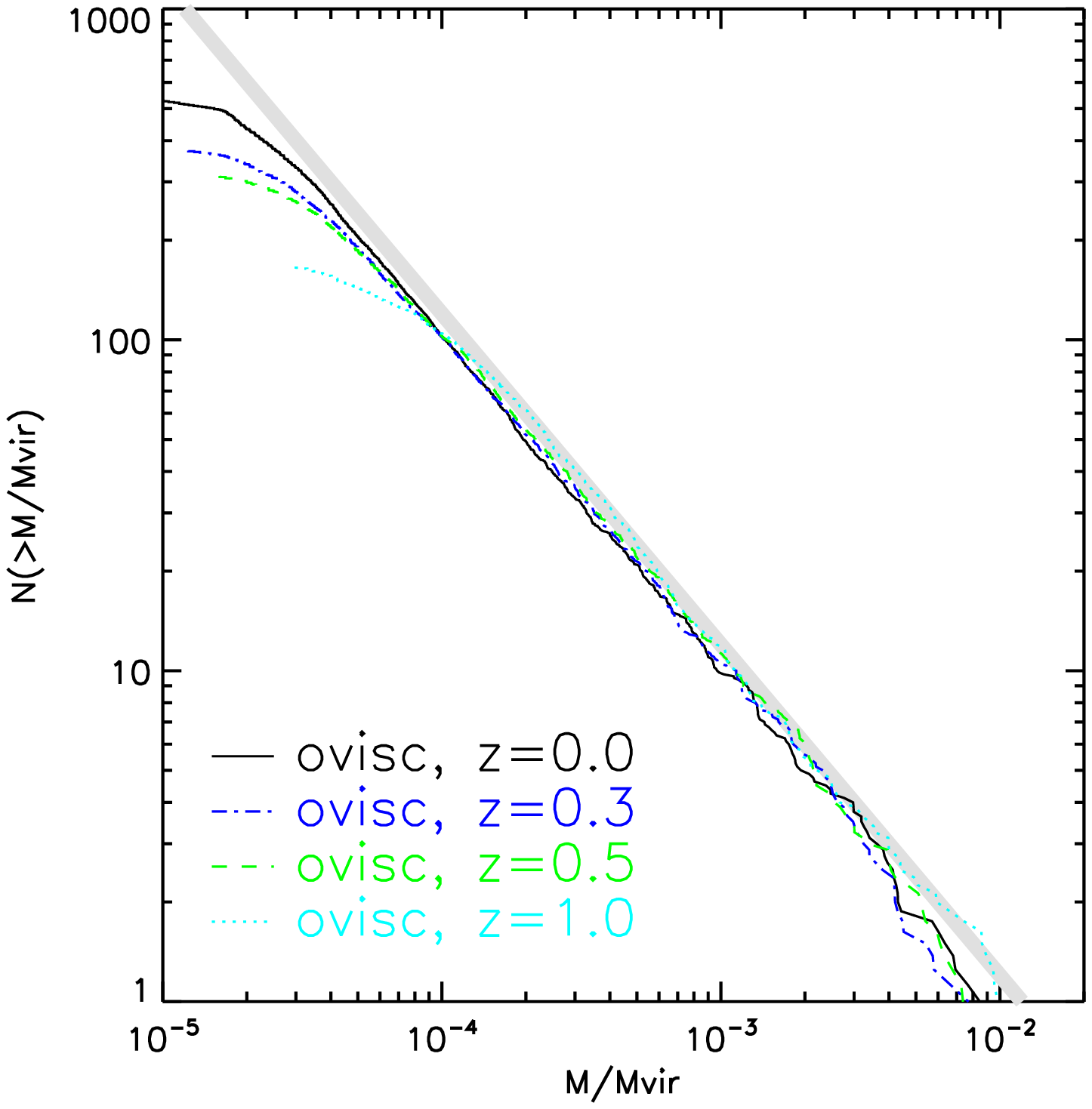}
\includegraphics[width=0.33\textwidth]{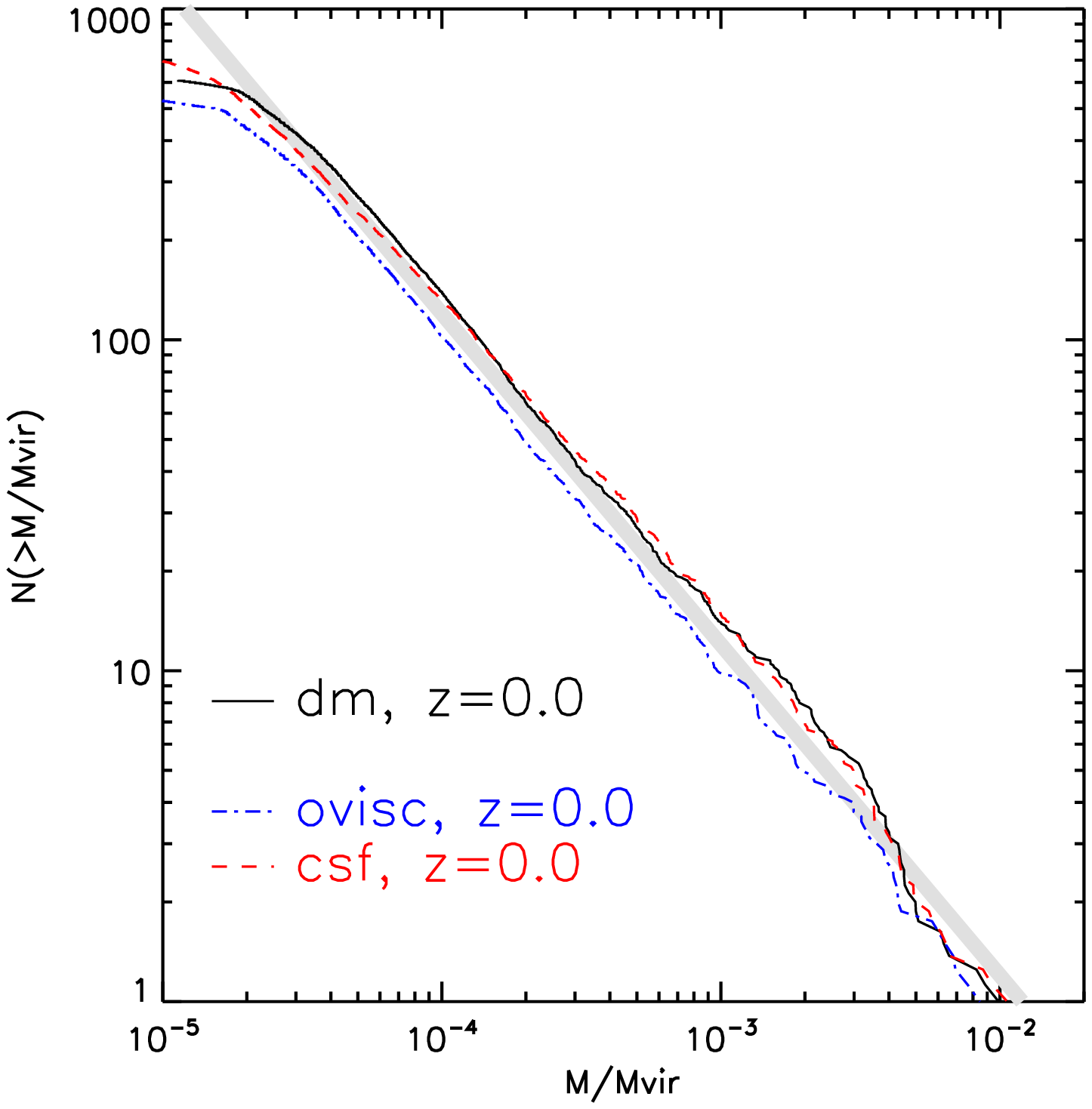}\\
\includegraphics[width=0.33\textwidth]{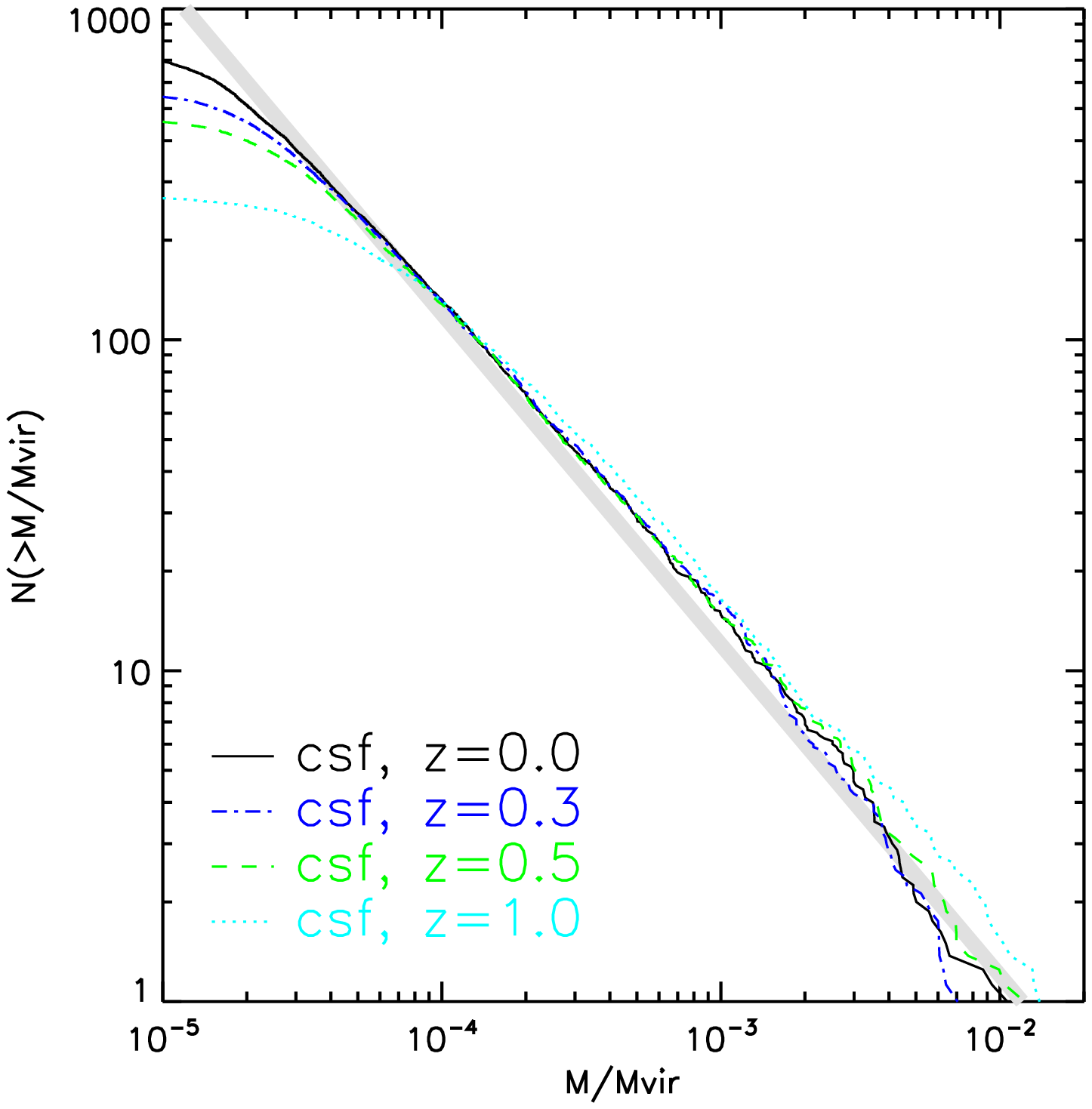}
\includegraphics[width=0.33\textwidth]{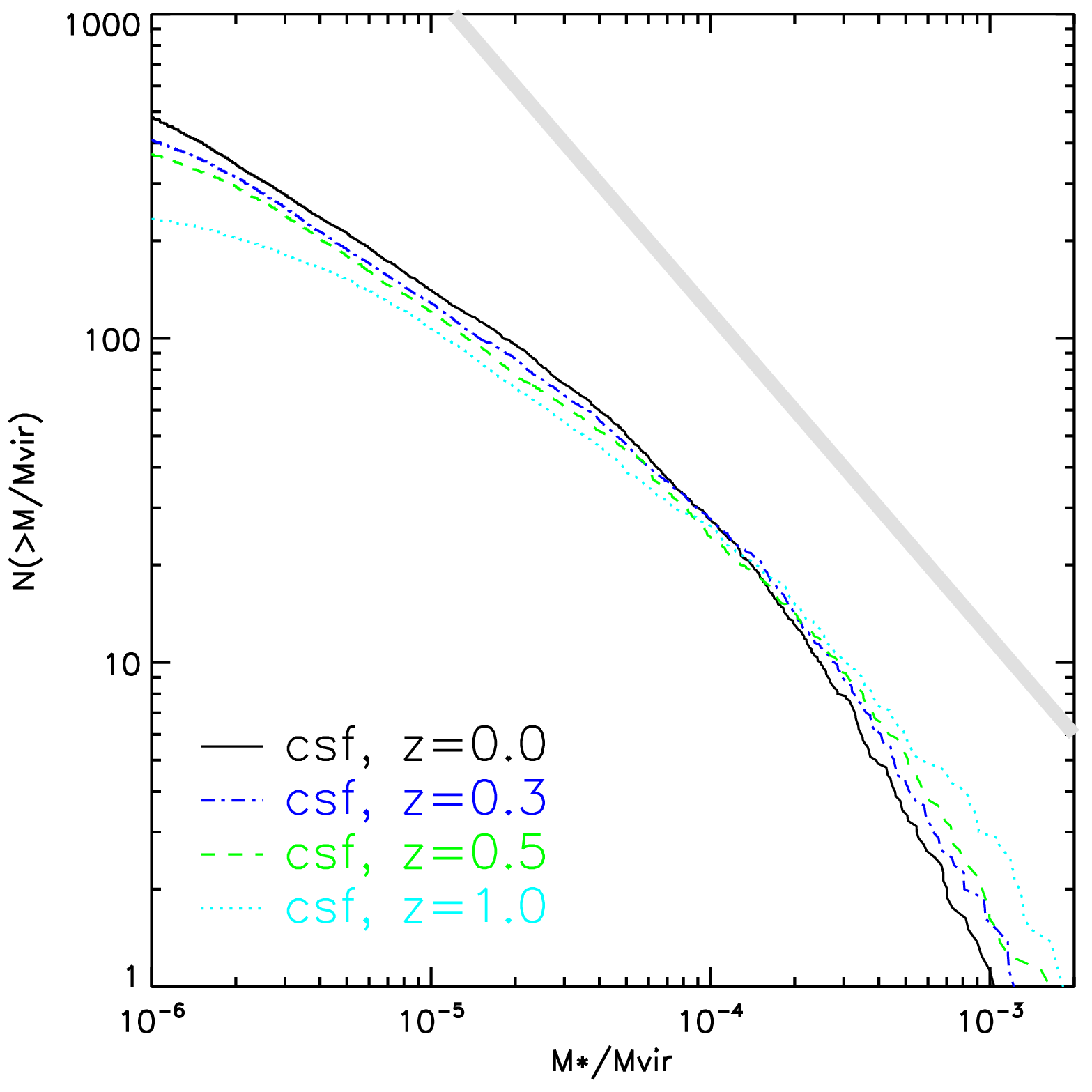}
\includegraphics[width=0.33\textwidth]{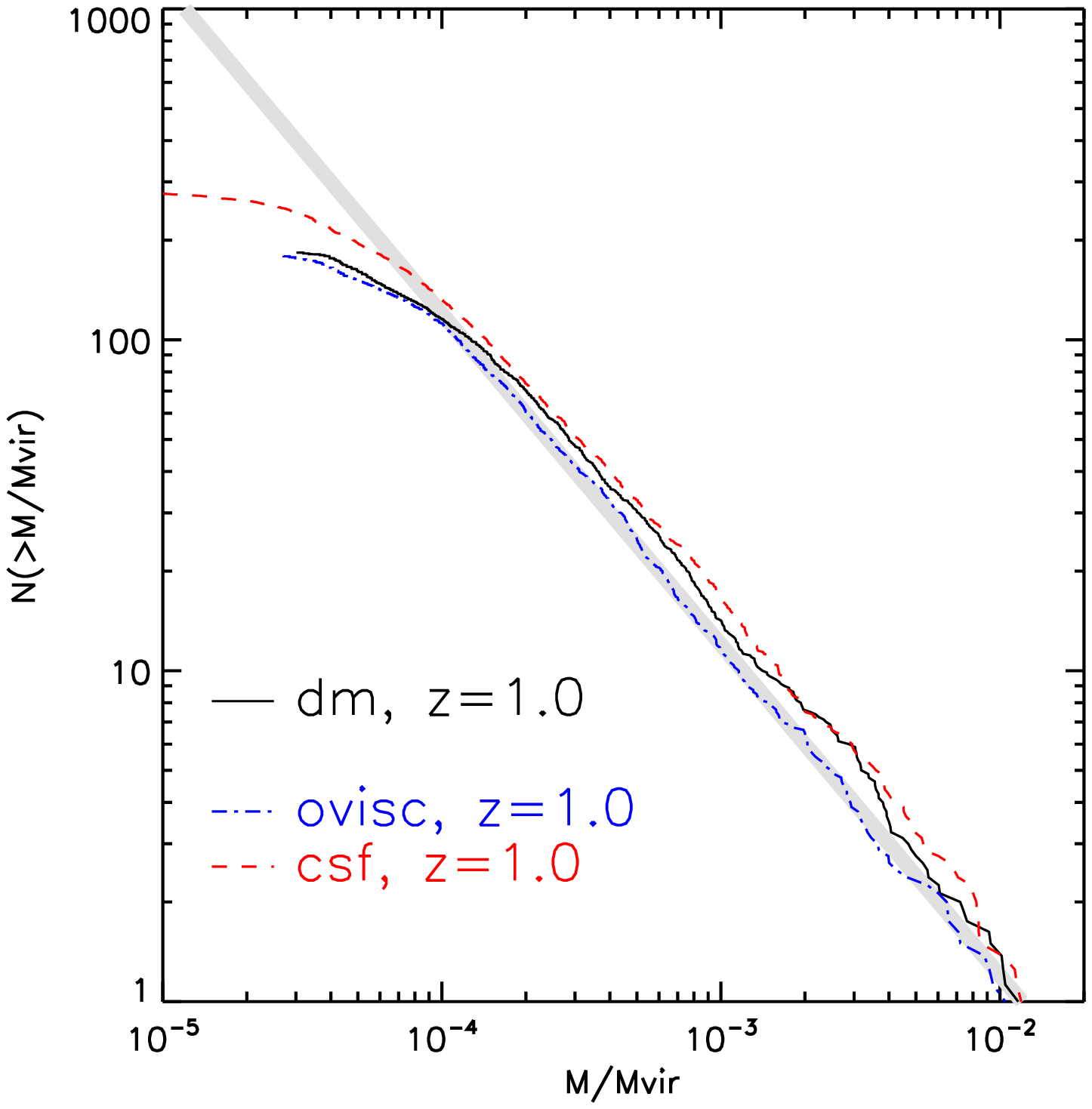}
\caption{Evolution of the subhalo mass functions for the different
  runs, averaged over the 8 clusters of the high mass sample. The two
  left and the two central panels show the mass functions at
  $z=0,0.3,0.5$ and $1$ with solid, dashed-dotted, dashed and dotted
  lines, respectively. Upper left, upper central and lower left panels
  are for total mass function for the DM, non-radiative {\em ovisc}
  and radiative {\em csf} runs, respectively. The lower central panel
  is for the stellar mass function of the radiative {\em csf} runs.
  The two right panels compare the total mass function for DM, {\em
    ovisc} and {\em csf} runs at $z=0$ (upper right panel) and at
  $z=1$ (lower right panel).  The thick grey lines in all panels mark
  the slope $-1$.} \label{fig:fig_evol}
\end{figure*}

Figure~\ref{fig:fig_phys} shows a comparison of the cumulative subhalo
mass functions for the different runs we carried out for the high-mass
{\it g51.a} cluster.  The left panel compares the total subhalo mass
function for the DM-only run with three non-radiative hydrodynamic
runs (solid lines) that used different treatments for the numerical
viscosity (see Section~\ref{sec:sim_phys} for details). Clearly,
introducing non--radiative hydrodynamics causes a decrease of the
total subhalo mass function by a sizable amount, and this is almost
independent of the detailed scheme used for the artificial
viscosity. If interpreted in terms of a shift in mass, the difference
between the subhalo mass in the DM-only and in the gas runs  can be
explained mostly by the  high efficiency of gas removal during the
infall of subhalos into the cluster. To demonstrate this, we increased
the mass of each sub-halo identified in the non-radiative run by an
amount corresponding to the cosmic baryon fraction. If gas is
completely stripped and this represents the only reason for the lower
mass function, then this correction should provide a mass function
consistent with that of the corresponding DM run.  However, the
stripped mass can not completely compensate for the mass difference
between the dark matter and the non-radiative runs.  Besides removing
gas, also their orbits are changed due to the effect of gas pressure
\citep[see also][]{2004MNRAS.350.1397T,2005AA...442..405P,Saro08},
also with some indications that the remaining dark matter sub-halos
might become easier to disrupt.

The central panel of Figure~\ref{fig:fig_phys} shows the cumulative subhalo
mass functions for the same cluster, but here comparing the DM run with the
different radiative runs, which differ in the efficiency of the kinetic
feedback (i.e. {\it csfnw}, {\it csf}, and {\it csfsw}), or in the presence of
thermal conduction (i.e., {\it csfc}).  The bulk of gas cooling and subsequent
star formation within subhalos takes place before the galaxies are falling
into the high--pressure environment of the ICM. Cooling has the effect of
concentrating cold gas and therefore also collisionless stars forming out of
the gas at the centres of these subhalos. This increases the concentration of
the mass distribution, and protects part of the baryonic mass from
ram--pressure stripping. As a result, the subhalo mass functions in these
radiative runs increase when compared to the non--radiative cases and become
quite similar to those of the DM-only runs, with only a marginal sensitivity to
the adopted intensity of the feedback.  Only in the run without any
kinetic feedback ({\it csfnw}), the resulting total subhalo masses are even
slightly larger than for the DM runs. This is probably the consequence of the
adiabatic contraction of the DM halos, in response to the rather strong
cooling taking place when galactic winds are turned off
\citep[e.g.,][]{2004ApJ...616...16G}.

It is worth mentioning that the change of the disruption of sub-halos
in the hydrodynamical simulations is due to a complex interplay
between different processes. Firstly, we find that the presence of
baryons makes the underlying dark matter distribution more
concentrated (see also \citealt{rasia2004,2006ApJ...651..636L}).  A
further increase in the concentration of the dark matter profiles is
also contributed by including radiative cooling. These findings are
reported in the left panel of Figure~\ref{fig:fig_new}, which shows
the concentration parameter of dark matter profiles as a function of
halo mass for all main halos. The results for the dark matter ({\it
  dm}) runs are in good agreement with the fit originally presented by
\citet{2004A&A...416..853D}. The concentrations in the non-radiative
runs and in the runs with cooling and star-formation increase on
average by 3-8\% and by 10-25\% with respect to the {\it dm } runs,
respectively. The expectation based on this result is then that the
halos in the non-radiative runs are more difficult to
destroy. However, as soon as the gas component is stripped from the
sub-halos, the latter react to this mass loss with a slight
expansion. In the non-radiative runs ({\it ovisc}) this expansion even
slightly over-compensates the increase of the concentration.  This is
demonstrated by the right panel of Figure~\ref{fig:fig_new}, which
shows the half-mass radii of all substructures found in the main
haloes at $z=0$ as a function of their total mass. We find that this
radius in the non-radiative runs ({\it ovisc}) is $\approx 2$\% larger
than in the pure {\it dm} runs. However, in the runs with cooling and
star formation ({\it csf}), the sub-halos remain more concentrated and
we find the half-mass radius to be $\approx 5$\% smaller than in the
{\it dm} run. This increase in halo concentration is sufficient to
compensate for the change of orbits due to the drag induced by the ram
pressure \citep[see][]
{2004MNRAS.350.1397T,2005AA...442..405P,Saro08}. As a consequence,
sub-haloes survive for a longer time in the {\it csf} runs, whereas
these effects are nearly compensating in the non-radiative runs with a
possible tendency for sub-haloes to be more easily disrupted when
compared to the {\it dm} only runs.

For the radiative runs, we note that a small but sizable population of small
subhalos exist that are dominated by star particles and whose DM component was
lost during the subhalo's evolution. These surviving compact cores of star
particles are still recognised by {\small SUBFIND} as self-bound structures,
and give rise to the low--mass extension of the subhalo mass functions. The
flattening of the mass function in these regimes shows that the number of
DM--poor galaxies is affected by the finite numerical resolution.

The right panel of Figure~\ref{fig:fig_phys} shows the subhalo mass functions
of just the stellar component, for the different radiative runs.  As expected,
the effect of changing the feedback strength is now clearly visible and can
lead to nearly a factor of ten in stellar mass between simulations without
kinetic feedback ({\it csfnw}) and with strong kinetic feedback ({\it csfsw};
see also \citealt{2006MNRAS.373..397S}). This comparison highlights the
existing interplay between the feedback, which regulates the star-formation
within subhalos and its progenitor halos, and the stripping due to the highly
pressurised environment of the intra--cluster medium. Although thermal
conduction is expected to alter the efficiency of gas stripping
\citep[e.g.,][]{2007arXiv0709.2772P}, its low efficiency in the
low--temperature subhalos results only in a marginal impact on the resulting
stellar mass function. In Appendix~\ref{sec:num}, we will further examine the
numerical convergence of the stellar mass function of the cluster galaxies.

\begin{figure*}
\includegraphics[width=0.33\textwidth]{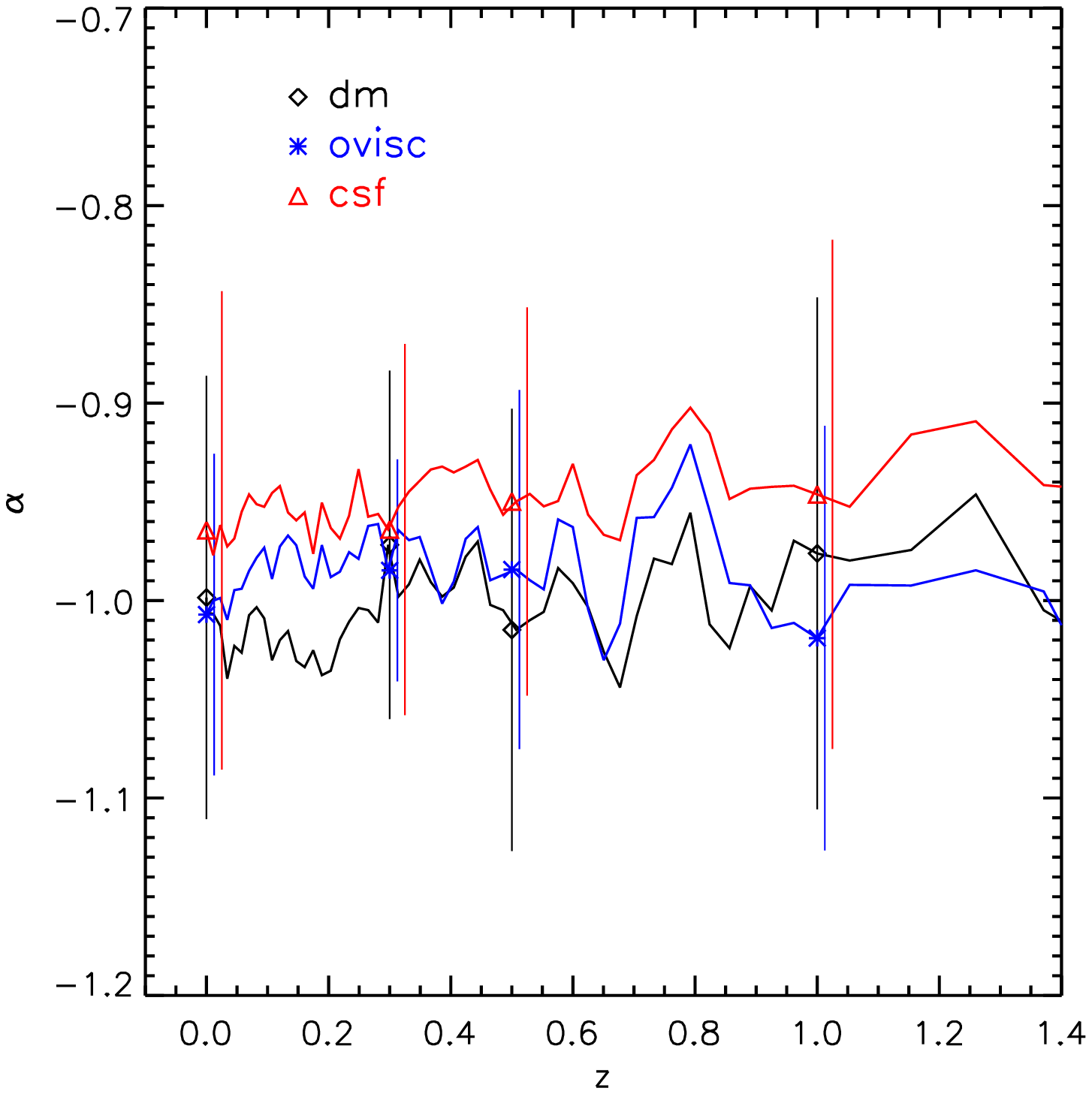}
\includegraphics[width=0.33\textwidth]{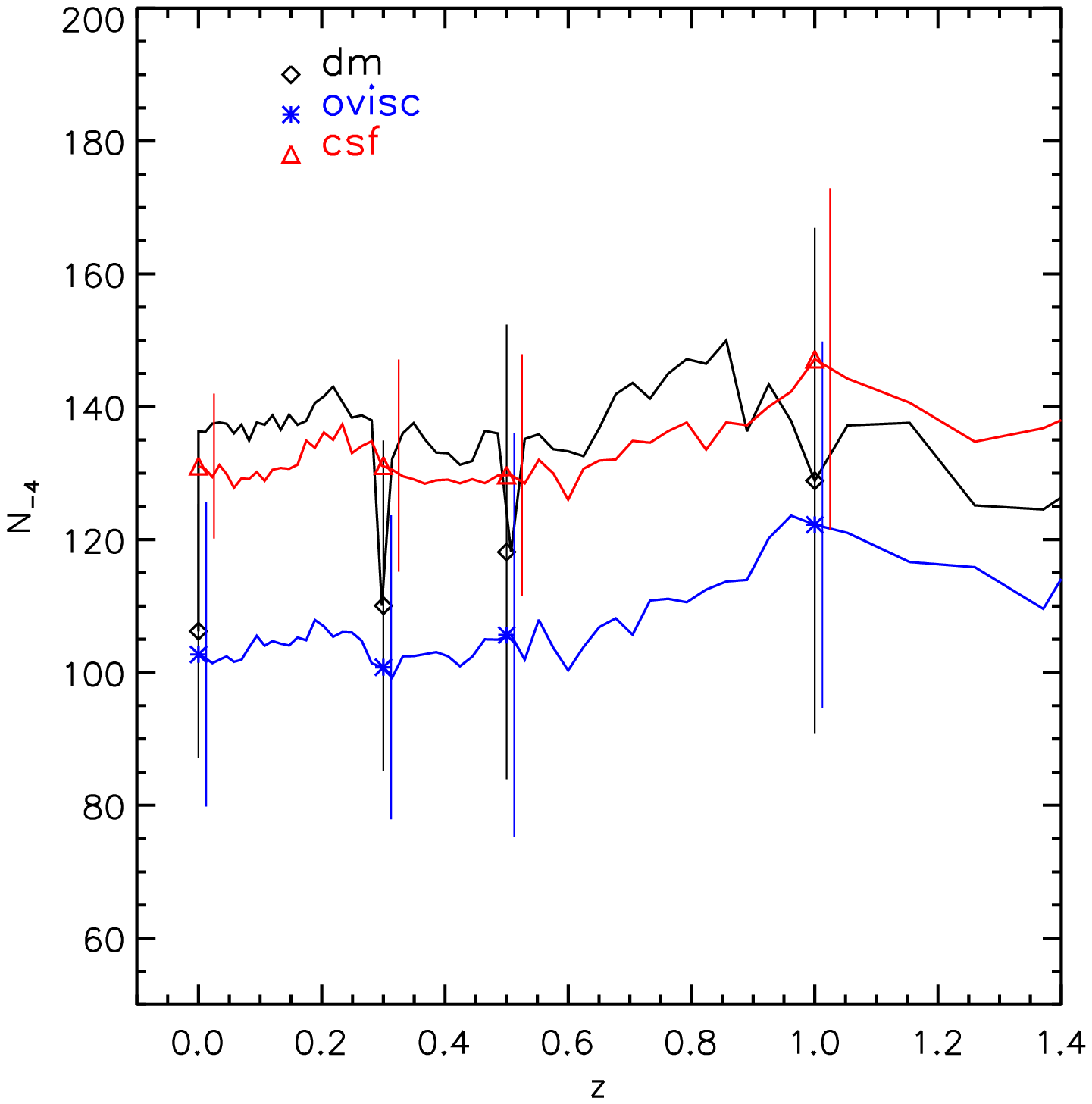}
\caption{The evolution of the parameters defining the power--law shape
  of the subhalo mass function (see equation \ref{eqn:fit}) for the
  progenitors of the 8 clusters belonging to the high--mass
  subsample. All the 52 outputs of the simulations out to $z=1.4$ have
  been used to produce the continuous lines.  Different
  colours/symbols are for the DM runs, for the ({\it ovisc})
  non--radiative run and for the ({\it csf}) radiative runs. The left
  and the right panels show the evolution of slope and normalisation,
  respectively. The error-bars mark the {\em rms} scatter computed over
  the 8 clusters of the high--mass subsample. For reasons of clarity,
  they are shown along with the symbols only at $z=0,0.3,0.5$ and 1.0,
  and are slightly displaced in the horizontal
  direction.} \label{fig:fit_param}
\end{figure*}

In Figure~\ref{fig:fig_evol}, we consider the time evolution of
the subhalo mass functions by showing averages for the main
progenitors of the eight high--mass clusters at different
redshifts. For the DM-only simulations we do not see any
significant evolution over the considered redshift range. This is
not directly in contrast to findings by
\citet{2004MNRAS.352L...1G}, as they define the virial mass with
respect to 200 times the critical density. If we stick to the same
definition of virial mass and radius we also get a evolutionary
trend in the same direction than found in
\citet{2004MNRAS.352L...1G}. The analysis of the average subhalo
mass function over the full sample of massive clusters confirms
the effect of the gas stripping in the non--radiative runs,
already discussed for the {\em g51} cluster at $z=0$, and the
counteracting effects by cooling and star-formation in the
radiative runs.

Furthermore, the rightmost upper and lower panels show the results at $z=0$
and $z=1$, respectively.  At intermediate subhalo mass (between $10^{11}{\rm
  M}_\odot$ and $10^{12}{\rm M}_\odot$), the effects of cooling and star
formation are that of slightly increasing the subhalo mass function compared
with the pure DM run. This is especially true at high redshift, when stripping
is less efficient. There are indications that, at the high mass end, stripping
can be still efficient enough to suppress the subhalo mass function, also in
the runs with cooling and star formation.  Again, a measurable effect is seen
from star dominated substructures for the low mass subhalos, indicating that
clumps of star particles, being more compact than DM subhalos, are more
resistant against tidal destruction. However, even the radiative simulations
do not predict that a significant fraction of galaxies without dark matter
halos survive down to $z=0$.

Figure~\ref{fig:fit_param} shows the evolution of the parameters
of the power-law fit to the total subhalo mass function, where we
have as usual restricted the fitting range to halos with at least
100 dark matter particles and $m_{\rm sub}/M_{\rm vir} < 0.01$,
which gives a reasonable fit over the whole considered redshift
range.  To guarantee robust fits, we here excluded those halos for
which there are less than 20 data-points in the estimate of the
cumulative mass function over this range. The slope of the total
subhalo mass function does not show a significant evolution in all
the runs. As for the normalisation (right panel of
Fig.~\ref{fig:fit_param}), its value for the non-radiative ({\it
ovisc}) gas run is always below that of the DM and of the
radiative {\it csf} gas runs.  There is also an indication that
the radiative run at high redshift has a normalization slightly
higher than the DM run. The stripping of the gas in the
non-radiative run is relatively independent of the halo mass, thus
leading to a slope similar to that of the DM case. We note that
there is a tentative indication that the condensation of baryonic
matter at the centre of the substructures in radiative runs leads
to a small systematic trend with redshift, leading to a slightly
flatter subhalo mass function at early redshift in the {\it csf}
runs.

\subsection{The composition of subhalos} \label{sec:sub_components}

In DM-only simulations, it is known that tidal stripping induces strong radial
dependences in the properties of substructures
\citep[e.g.][]{2004MNRAS.352L...1G,2004MNRAS.348..333D}.  For example, the
substructure abundance is antibiased relative to the mass distribution of the
cluster, and the average concentration of subhalos increases towards the
cluster centre.  The efficiency of gas stripping from substructures also
depends strongly on cluster-centric distance
\citep[see][]{1972ApJ...176....1G,2008MNRAS.383..593M}, hence we expect radial
variations in the relative content of baryons in subhalos, which we analyse in
this section.

The problem is made especially interesting by the complex non-linear interplay
between the condensation of baryons due to radiative cooling, the
gas--dynamical stripping processes, and the resulting orbital evolution of
affected subhalos.  Indeed, while gas stripping in the high--pressure ICM
makes subhalos more fragile, the formation of a compact core of cooled gas and
stars makes subhalos more resistant to disruption. As a result, we expect that
the relative stellar content of a subhalo and its capability to retain a
gaseous halo will depend both on the details of the physics included in the
simulation and on its cluster-centric distance.

\begin{figure}
\begin{center}
\includegraphics[width=0.33\textwidth]{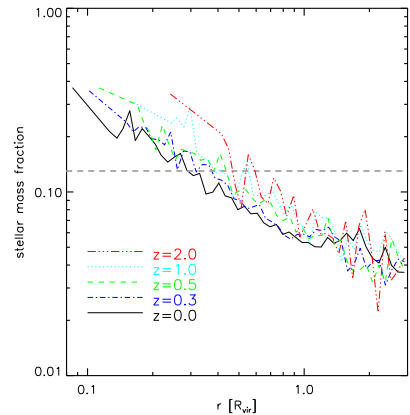}
\end{center}
\caption{The evolution of the radial profiles of the stellar mass
  fraction within subhalos for the radiative {\em csf} run. Each curve
  is obtained by averaging over the 8 clusters belonging to the
  high--mass sample. Only subhalos with masses $\le 2\times10^{12}
  {\rm M}_\odot$ have been included in this analysis.  The horizontal dashed
  line marks the cosmic baryon fraction assumed in our simulations.}
\label{fig:comp_stars}
\end{figure}

In order to investigate this in more detail, we show in Figure
\ref{fig:comp_stars} the radial dependence of the fraction of subhalo
masses contributed by stars for the radiative runs including the
reference kinetic feedback ({\em csf}), at three different epochs
($z=0,1,2$ from left to right).  In each panel, the different lines
indicate different limits for the total mass of the selected
subhalos. Quite clearly, the stellar component becomes progressively
more important in subhalos found at a smaller cluster-centric
distance. This is consistent with the expectation that the strong
tidal interactions in the cluster central regions are efficient in
removing the dark matter component of the subhalos, while the
gravitationally more tightly bound clumps of star particles are more
resistant.  Fig.~\ref{fig:comp_stars} also shows that the radial
dependence of the subhalo mass fraction in stars extends well beyond
the clusters' virial radius $R_{\rm vir}$, thus indicating that the
cluster environment affects the properties of the galaxy stellar
population over a region much larger than the virial one \citep[see
  also][]{2006MNRAS.373..397S}.  Note that a substantial fraction of
these subhalos detected at $r>R_{vir}$ may have already crossed the
cluster virial region in earlier passages and constitute the so-called
``backsplash population''
\citep[see][]{2005MNRAS.356.1327G,2008arXiv0801.1127L}. Their orbits
have brought them into the outer parts again, with a fraction of them
eventually destined to escape the cluster.

Towards the centre of the clusters, the star fraction within subhalos
reaches values higher than the cosmic baryon fraction.  At first
glance, this result lends support to the assumption made by many
semi--analytical models (SAMs) of galaxy formation that galaxies can
preserve their identity for a while even after their host DM--halos is
destroyed by the tidal interaction within clusters
\citep{2004MNRAS.352L...1G}.  However, in semi--analytical models
\citep[like][]{2007MNRAS.375....2D} based on DM simulations of
comparable resolution \citep[like the {\it Millennium
    Run,}][]{2005Natur.435..629S} about half of the galaxies within
the virial radius of the parent cluster are expected to have lost
their DM halos at $z=0$, while in our hydrodynamical simulations we
find that only $\approx 20$ per cent of the galaxies have a star
component which exceeds the mean cosmic baryon fraction (i.e. $m_{\rm
  sub}^* > 0.13\, m_{\rm sub}$), while only $2$ per cent of the
galaxies have their mass dominated by star particles (i.e.,
$m_\mathrm{sub}^* > 0.5 m_\mathrm{sub}$).  This implies that only a
small fraction of our star dominated systems can be classified as {\it
  orphan} galaxies, which are defined as those galaxies not having a
counterpart in the corresponding dark matter only run.  A precise
comparison requires however a carefully matched magnitude selection,
which is outside the scope of the present paper. Future, yet higher
resolution hydrodynamical simulations should in principle allow an
accurate calibration of the SAM assumptions about the relative
survival time of satellite galaxies once their parent dark matter halo
cannot be tracked any more.

\begin{figure*}
\includegraphics[width=0.33\textwidth]{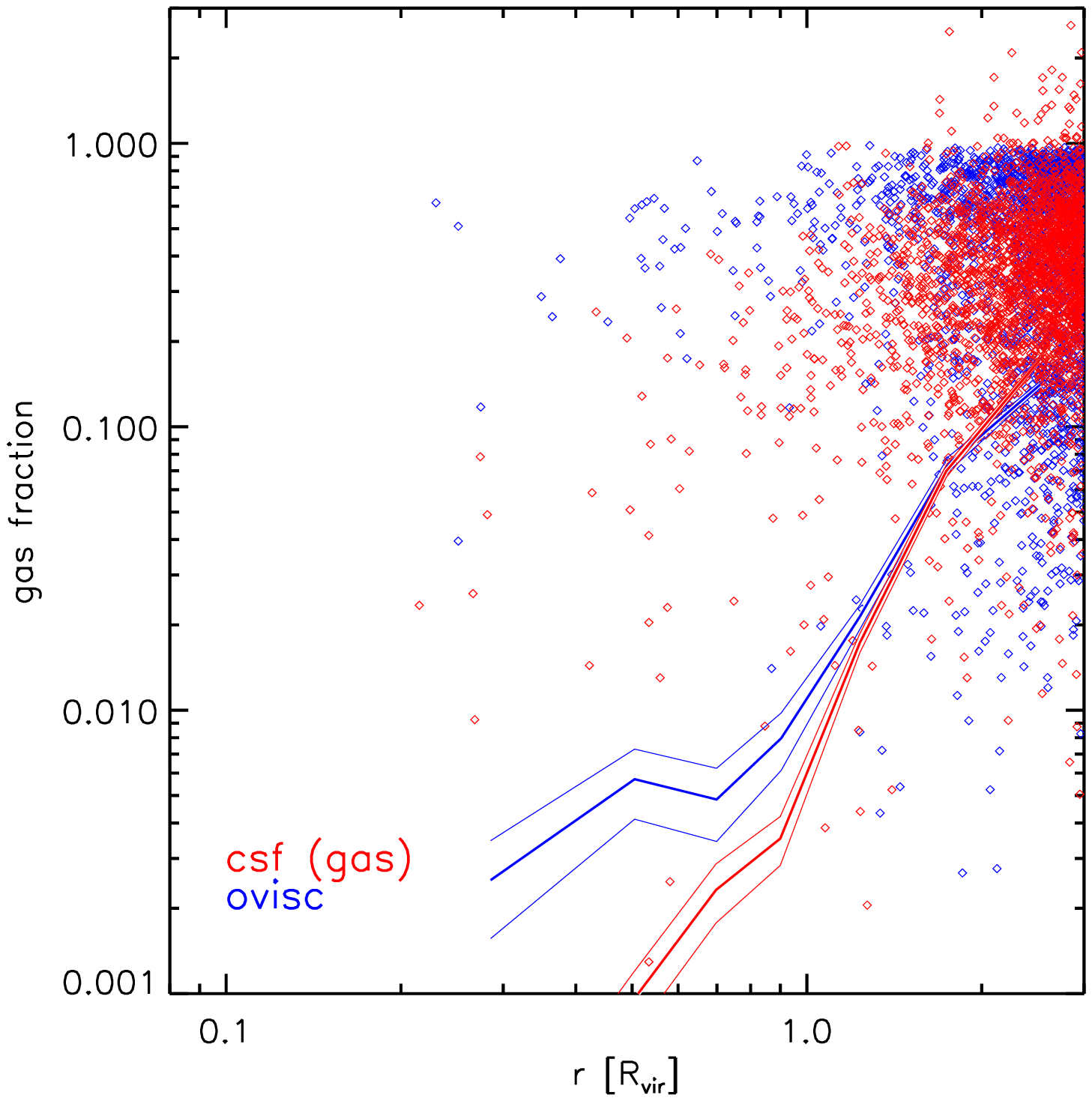}
\includegraphics[width=0.33\textwidth]{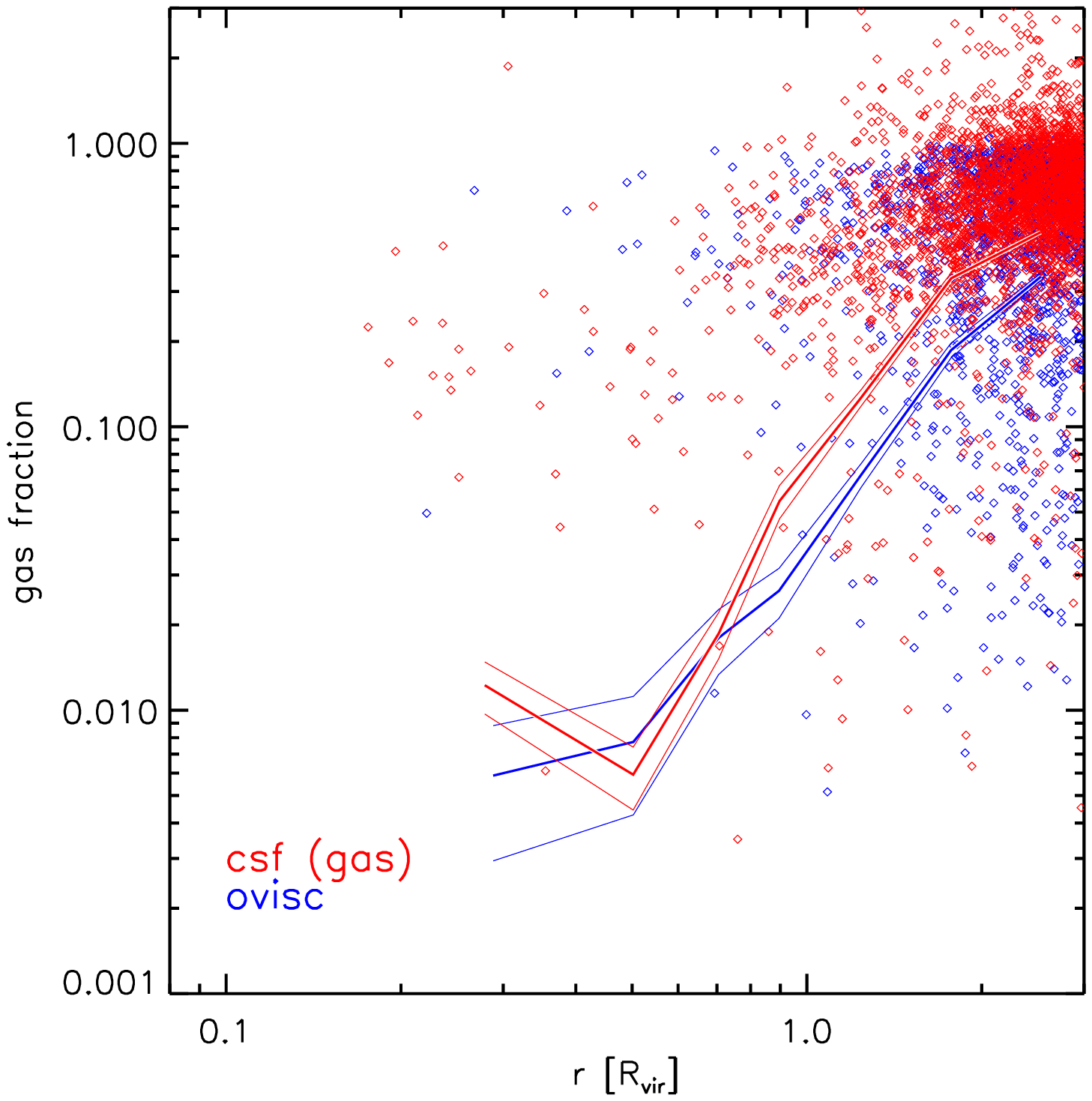}
\includegraphics[width=0.33\textwidth]{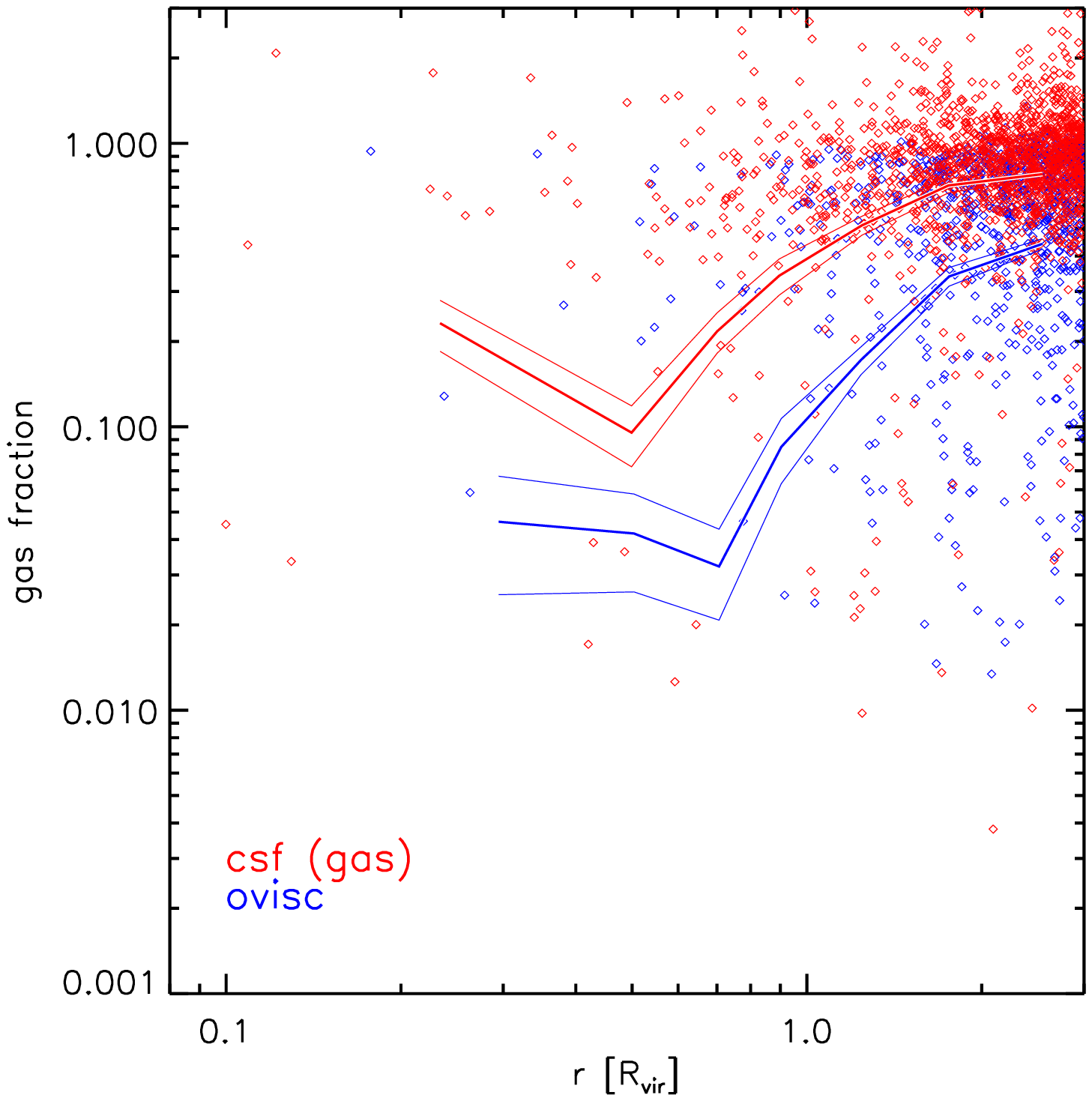}
\caption{The radial dependence of the fraction of subhalo mass in gas,
  in units of the cosmic baryon fraction, for the eight clusters
  belonging to our high-mass sample. Black and blue symbols are for the
  subhalos retaining a gravitationally bound gas component for the
  non--radiative {\em ovisc} and for the radiative {\em csf} runs,
  respectively. The heavy continuous lines show the radial dependence
  of the gas fraction after averaging over all subhalos, including
  those not retaining any gas, while the light lines indicate the
  uncertainties arising from the small number statistics of the
  involved subhalos. Left, central and right panels are for $z=0,1$
  and 2, respectively.}
\label{fig:comp_gas}
\end{figure*}

A related question concerns the ability of infalling subhalos to
retain a gravitationally bound component of diffuse gas during the
cluster evolution. We show in Figure~\ref{fig:comp_gas} the
cluster-centric dependence of the subhalo mass fraction associated
with gravitationally bound gas (in units of the cosmic baryon
fraction) from the set of eight massive clusters, at $z=0$, $1$ and
$2$. In each panel, we compare results for the {\em ovisc}
non--radiative run with those for the radiative {\em csf}
run. Furthermore, the solid lines indicate the radial dependence of
the average subhalo gas mass fraction, which is computed by including
also those subhalos which do not have any gravitationally bound gas
component.  We indicate with the thin lines the Poisson uncertainties
in the estimate of the average gas mass fraction, by assuming them to
be proportional to the square-root of the number of galaxies in each
radial bin. At $z=0$ we note that, whenever a subhalo retains a gas
component, the associated mass fraction is higher in the
non--radiative run than in the radiative one. This is related to the
fact that a significant fraction of this gas is converted into stars
in the radiative case.

On the other hand, the central concentration of star particles in
subhalos in the radiative runs makes them more resistant against
ram--pressure stripping.  As a consequence, a larger number of
subhalos preserve a gas component, thus explaining the higher average
gas fraction at all radii. Within the virial radius, we find at $z=0$
that about $\simeq 1$ per cent of the resolved subhalos still have
some self-bound gas, both in the {\it ovisc} and in the {\it csf}
runs, with a marginal indication for a smaller amount of gas in the
{\it csf} clusters, due to its conversion into stars. As expected, at
high redshift ram--pressure stripping has not yet been as effective,
thus explaining the larger fraction of self--bound gas, which is
$\simeq 4$ per cent and $\simeq 6$ per cent for the {\it ovisc} and
the {\it csf} runs, respectively.  Again, we note that the radial
dependence of the gas fraction extends well beyond the virial radius,
thus confirming that the overdense cluster environment affects the
structure and evolution of subhalos over a fairly large region
surrounding the clusters \citep[see also][]{2006MNRAS.370..656D}. Also
note that at very high redshift the collapse of the gas due to
efficient cooling inside the subhalos can even lead to gas fractions
larger than the cosmic values in the radiative runs.

An important question for hydrodynamical simulations of cosmic structure
formation is how faithfully they are able to describe the gas--dynamical
processes which determine the stripping of the gas in substructures moving in
a high density environment. In these processes, hydrodynamical fluid
instabilities that are difficult to resolve can play an important role.  In
general, the faithfulness of the simulation depends both on the hydrodynamical
scheme adopted (i.e., Eulerian or Lagrangian, implementation of gas viscosity,
etc.)  and on the numerical resolution achieved \citep[e.g.,
][]{2005MNRAS.364..753D,2006MNRAS.371.1025S,2007MNRAS.380..963A,2008arXiv0801.4729I}.
Although a detailed study of these numerical aspects is beyond the scope of
this paper, it is interesting to examine the impact of the adopted artificial
viscosity parameterization on the gas stripping from subhalos.

\begin{figure*}
\includegraphics[width=0.33\textwidth]{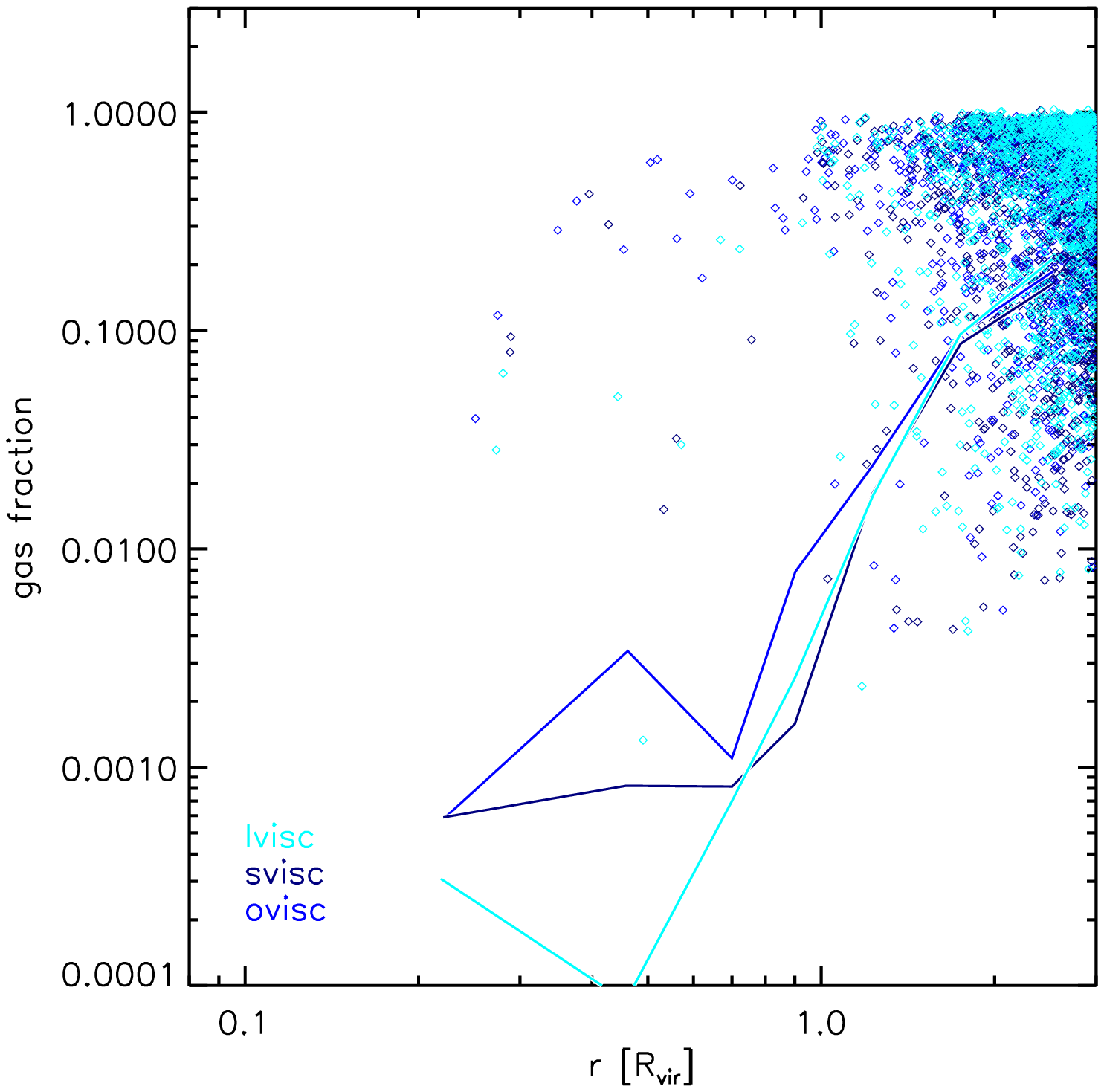}
\includegraphics[width=0.33\textwidth]{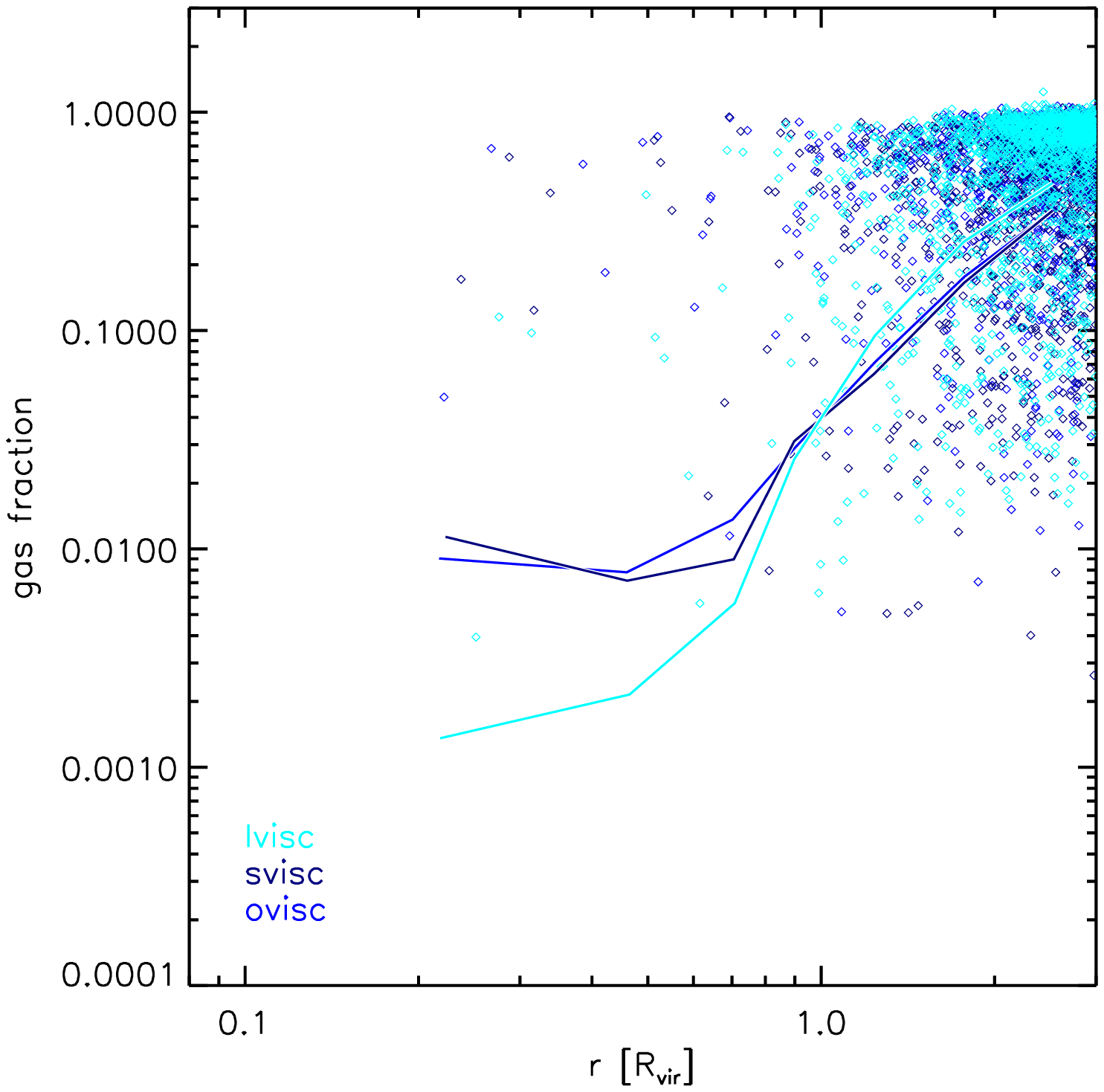}
\includegraphics[width=0.33\textwidth]{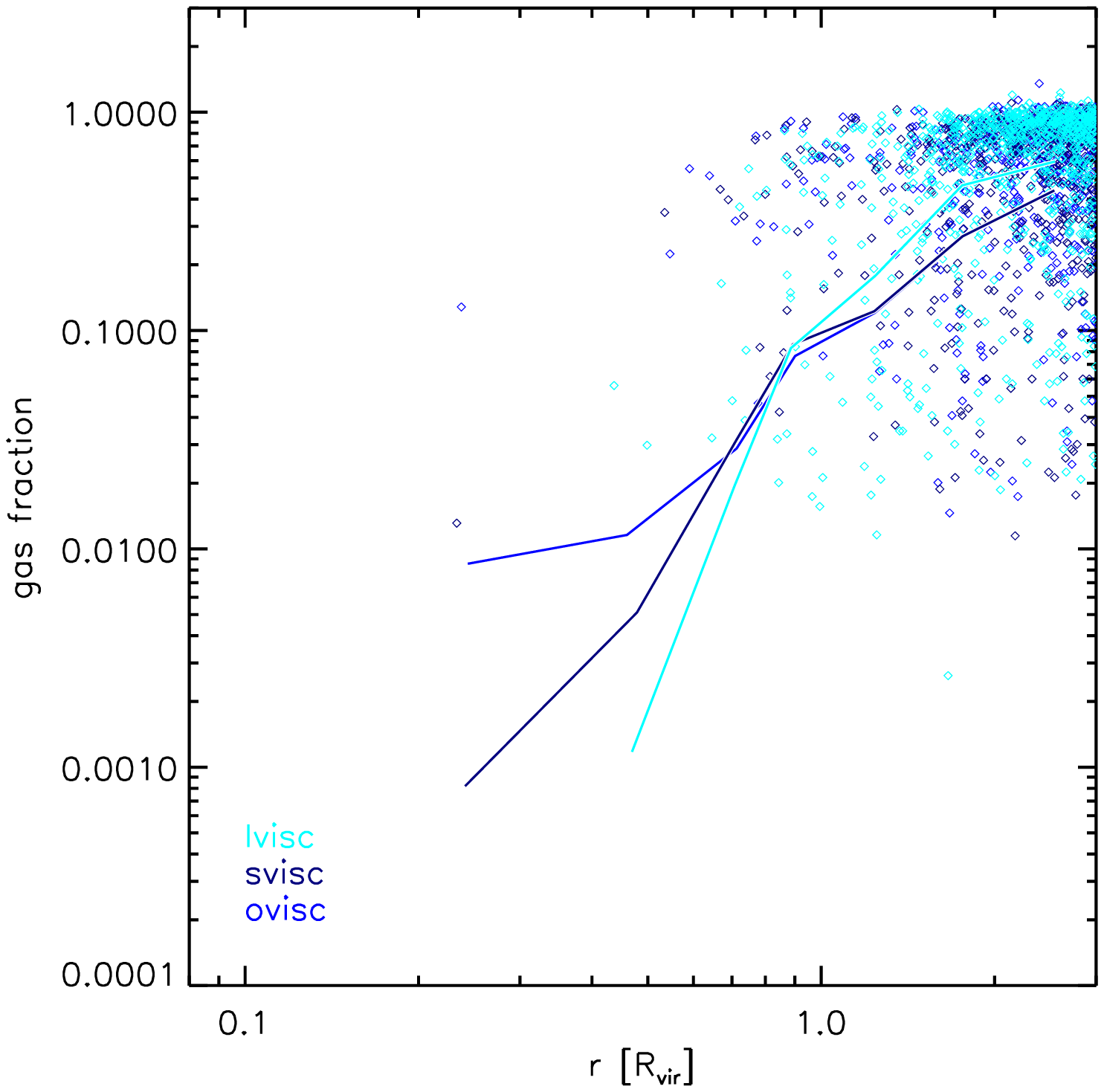}
\caption{The same as in Figure \ref{fig:comp_gas}, but comparing
  results for the different implementation of artificial
  viscosity. Results are shown for the four clusters of the high--mass
  sample for which runs with the different viscosity schemes have been
  carried out.}
\label{fig:comp_visc}
\end{figure*}

Figure~\ref{fig:comp_visc} shows the subhalo gas mass fraction as a
function of cluster-centric distance for the four clusters of our
high-mass sample, for which runs with different artificial viscosities
have been carried out (i.e., simulations {\it g1.a}, {\it g8.a}, {\it
  g51.a} and {\it g72.a}; see Table \ref{tab:cluster}.  Similarly to
the results obtained by \citet{2005MNRAS.364..753D} for the predicted
ICM turbulence, the {\it ovisc} and {\it svisc} implementations
give quite similar results, whereas the {\it lvisc} low--viscosity
scheme leads to smaller gas-mass fractions inside the halos and
slightly higher gas-mass fractions outside the virial radius. This is
a clear indication that numerical viscosity has a measurable effect on
the stripping of gas from the subhalos, as the latter depends on the
environment.

The different behaviour of the lower--viscosity scheme can be
understood by recalling that this approach provides less efficient
viscous stripping, which explains the larger gas fraction associated
with subhalos found in the cluster outskirts. On the other hand, as
sub-halos cross the virial radius and enter a progressively more
pressurised environment, they should experience hydrodynamical
instabilities in the shear-flows around the subhalo's gas, such as the
Kelvin--Helmholtz instability, which may be poorly represented in SPH
\citep{2007MNRAS.380..963A}. While the low-viscosity scheme does not
improve the SPH description to a point where these instabilities are
captured satisfactorily with SPH, it still improves the description of
shear flows, so that the subhalo gas can be dissolved and dispersed in
the hotter atmosphere over a relatively short time-scale.

\section{Conclusions} \label{sec:conc}

In this study, we presented an analysis of the basic substructure
properties in high--resolution hydrodynamical simulations of
galaxy clusters carried out with the TreePM-SPH code {\small
GADGET}. We used a fairly large set of 25 clusters, with virial
masses in the range $1-33 \times 10^{14}\msun$, out of which eight
clusters have masses above $10^{15}\msun$. The identification of
the substructures has been carried out with the {\small SUBFIND}
algorithm \citep{2001MNRAS.328..726S}, which we suitably modified
to account for the presence of gas and star particles in
hydrodynamical simulations. The primary aim of our analysis was to
quantify the impact that gas physics has on the properties of
subhalos, especially on their abundance. Note that previous work
has so far almost exclusively analyzed substructures in dark
matter-only simulations, so that our study is one of the first
attempts to directly compare the DM-only results to those of
hydrodynamical simulations that also account for baryonic physics.

We analyzed non--radiative hydrodynamical simulations as well as
radiative simulations, including star formation and different
efficiencies for energy feedback associated with galactic outflows.
Also, we examined the sensitivity of our results with respect to
different parameterizations for the artificial viscosity needed in
SPH. The main results of our analysis can be summarized as follows.

\begin{itemize}
\item Consistent with previous studies
  \citep{2001MNRAS.328..726S,2004MNRAS.348..333D}, we find that the
  subhalo mass function in DM-only runs converges well for different
  numerical resolution over the mass range where substructures are
  resolved with at least $\simeq 30$ gravitationally bound DM
  particles. We extend this result to show that the same convergence
  also holds well for the total subhalo mass function in
  hydrodynamical runs, irrespective of whether cooling and star
  formation are included in the simulations.

\item In non--radiative hydrodynamical simulations, the presence of a
  gas component leads to a reduction of the total subhalo mass
  function with respect to DM-only runs. On average, the reduction in
  mass of the substructures is slightly larger than expected from 
  the complete stripping of the baryonic component. This is due to a
  combination of the modification of the subhalo orbits arising from
  the pressure force exerted by the intra--cluster medium. Furthermore,  
  we see indications for an increased susceptibility of the surviving DM 
  halos to tidal disruption.

\item In general we find that only a very small fraction of subhalos,
  of order of one percent, within the cluster virial radii maintain a
  self-bound atmosphere of hot gas. This indicates a high efficiency
  for the stripping of the hot gas component in our simulations. The
  radial dependence of the mean gas fraction within subhalos indicates
  that gas stripping starts already at large cluster--centric
  distances, beyond the virial radius.  Using a scheme for reduced
  artificial viscosity has the effect of reducing viscous stripping in
  the cluster periphery, while increasing the stripping within the
  virialized high-density region.

\item In radiative runs that include star formation, some of the
  baryons are protected from stripping due to their concentration at
  the centres of the subhalos.  In this case, substructures have
  masses which are comparable to, or are slightly larger than in the
  DM-only runs.  The stellar mass fraction is also found to strongly
  increase for subhalos close to the centre. This confirms that
  compact clumps of star particles are indeed more resistant against
  tidal destruction than DM subhalos. Despite this, we find that only
  a small fraction of subhalos are dominated by their stellar
  component. As a result, the number of star--dominated galaxies is
  smaller than expected when, like usually assumed in semi--analytical
  models of galaxy formation, they are allowed to survive for a
  dynamical friction time after the disruption of their DM halo. This
  suggests that our hydrodynamical simulations at their current
  resolution may not be able yet to accurately follow the survival of
  galaxies within clusters after the destruction of their DM subhalos.
\end{itemize}

Our study shows that the technical improvements we implemented in
{\small SUBFIND} allow a reliable identification of substructures
in hydrodynamical simulations that include radiative cooling and
star formation. This allows a direct identification of satellite
galaxies not only in clusters of galaxies, but also in future
high-resolution hydrodynamical simulations of galaxy-sized halos.
This is essential to make direct contact with observational data,
and also allows studies of the structural impact of baryonic
physics on substructure. As our result show here, gas physics {\em
does alter} substructure statistics in interesting ways, but the
effects depend on physical processes such as radiative cooling and
feedback. This also means that the approximate self-similarity of
substructure properties that has been found in DM-only simulations
of clusters and galaxy-size objects \citep{1999ApJ...524L..19M} is
not expected to hold nearly as well in simulations with radiative
cooling. For example, in galaxy-size halos, many subhalos will be
so small that they do not experience atomic cooling; as a result
their behaviour should be close to our non-radiative results, and
we would here expect a reduction of their abundance relative to
the DM-only case.

In future work, it will be important to further improve the numerical
description and resolution of our simulations, in order to be able to more
accurately follow in particular the star--dominated substructures (i.e.,
galaxies) within galaxy clusters. This will then also provide important input
for the semi--analytical models of galaxy formation, allowing a check and
calibration of their dynamical friction and gas stripping prescriptions.


\section*{acknowledgements}
We thank an anonymous referee for constructive comments which
helped to improve the presentation of the results.
The simulations were carried out on the IBM-SP4 machine at the
``Centro Interuniversitario del Nord-Est per il Calcolo Elettronico''
(CINECA, Bologna), with CPU time assigned under an INAF-CINECA grant,
on the IBM-SP3 at the Italian Centre of Excellence ``Science and
Applications of Advanced Computational Paradigms'', Padova and on the
IBM-SP4 machine at the ``Rechenzentrum der Max-Planck-Gesellschaft''
at the ``Max-Planck-Institut f\"ur Plasmaphysik'' with CPU time
assigned to the ``Max-Planck-Institut f\"ur Astrophysik''.
K.~D.~acknowledges the hospitality of the Department of Astronomy of
the University of Trieste and the receipt of a ``Short Visit Grant''
from the European Science Foundation (ESF) for the activity entitled
``Computational Astrophysics and Cosmology''. We thank Gerard Lemson
for all the support with the database and related issues. We also
thank Gabriella De Lucia and Alex Saro for a number of useful
discussions. This work has been partially supported by the PD51 INFN
Grant, by the ASI-COFIS and by the ASI-AAE contracts.

\bibliographystyle{mn2e}
\bibliography{master3}

\appendix


\section{Numerical stability} \label{sec:num}

This Appendix is devoted to a discussion of the numerical stability of
the results presented for the properties of the substructures in
simulated clusters. In the first part, we will discuss the robustness
of the results against the choice of the subhalo finding algorithm. In
the second part we will discuss the stability of the subhalos mass
functions against numerical resolution, choice of the minimum SPH
smoothing length and scheme to add gas particles to initial
conditions.

\begin{figure*}
\includegraphics[width=0.33\textwidth]{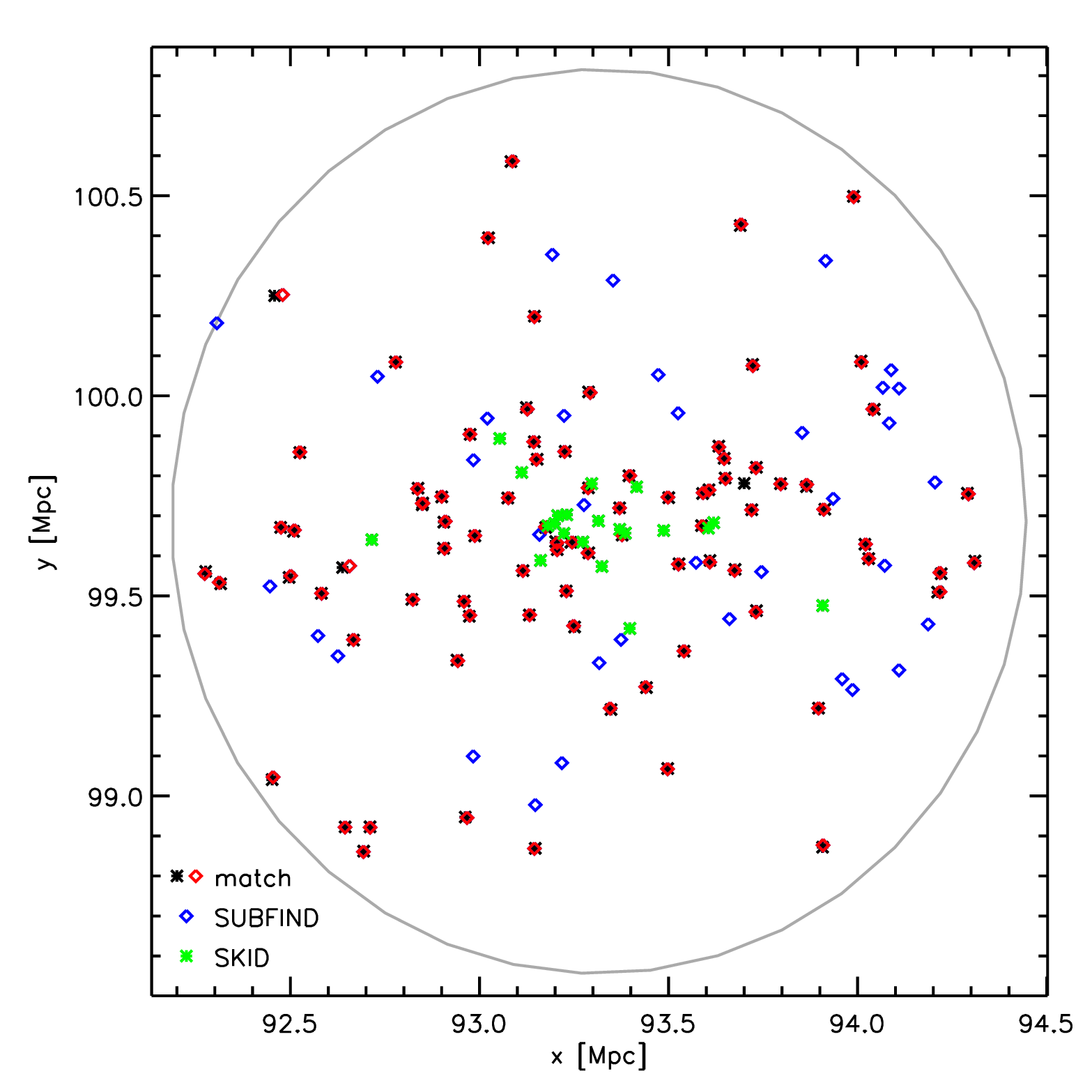}
\includegraphics[width=0.33\textwidth]{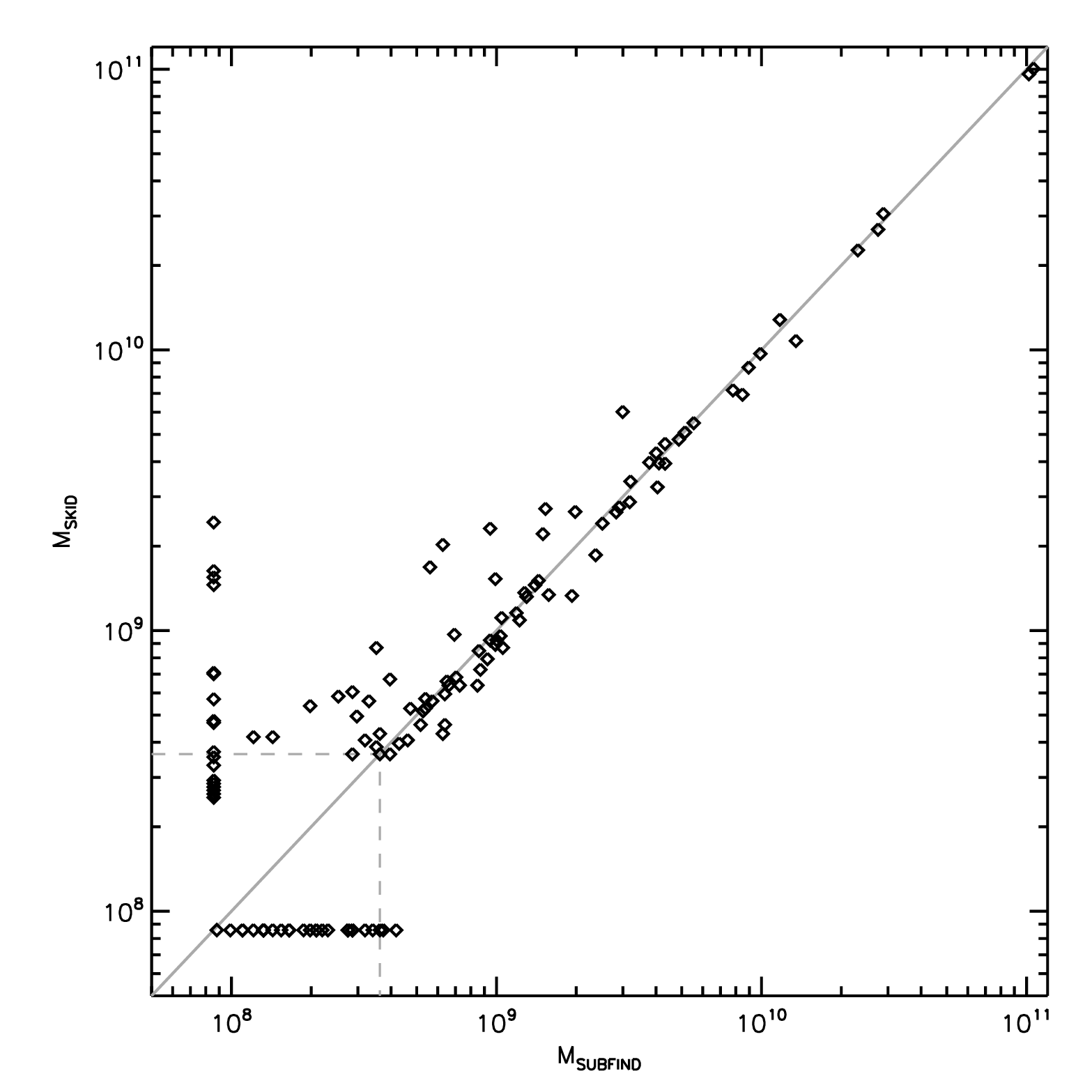}
\includegraphics[width=0.33\textwidth]{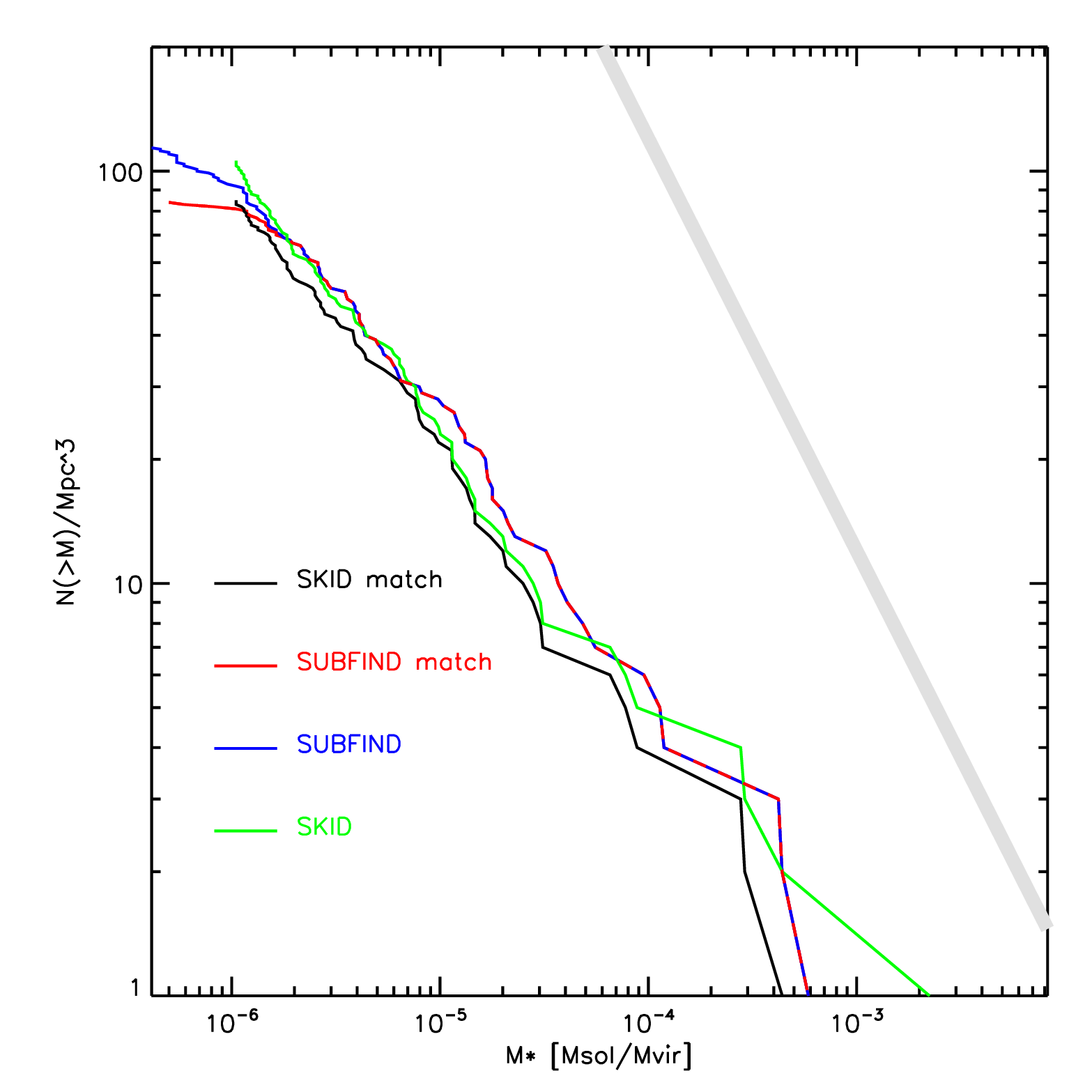}
\caption{Comparison of the galaxies detected by {\small SKID} and by
  {\small SUBFIND} for the R3 run at $z=0$.  The left panel shows the
  position of the individual substructures, with different
  symbols/colours for galaxies identified with the two algorithms, as
  specified in the labels. Red diamonds and black crosses indicate
  galaxies identified by {\small SUBFIND} and by {\small SKID},
  respectively, when the positions provided by the two algorithms
  differs by less than the half mass radius of the galaxy in the
  {\small SUBFIND} identification.  Blue diamonds and green crosses
  are for the galaxies identified by {\small SUBFIND} and by {\small
    SKID}, respectively, when they do not have a counterpart
  identified by the other algorithm. In the results from {\small
    SUBFIND}, we do not include identified substructures which do not
  have a stellar component. The middle panel shows a one-to-one
  comparison between the stellar masses assigned by the two algorithms
  to the identified galaxies.  The symbols lining up
  vertically/horizontally are those identified by one algorithm, but
  with no match in the galaxy sample from the other algorithm. The
  dashed lines mark the stellar mass limit applied to the {\small
    SKID} detection. The right panel shows the cumulative stellar
  subhalo mass function from the two identification schemes. We also
  plot the mass functions for the galaxies identified from one
  algorithm and having a match in the galaxy sample derived from the
  other algorithm.}
\label{fig:comp_ss}
\end{figure*}

\subsection{Substructure detection}

In Section~\ref{sec:detection} we described our modification of the
{\small SUBFIND} algorithm \citep{2001MNRAS.328..726S} to allow a
clear detection of substructures with various compositions in
hydrodynamical simulations including cooling and star formation.  To
verify our method, we reanalyzed some of the clusters studied in
\citet{2007MNRAS.377....2M} with the new method and compared the
identified stellar components of the individual galaxies. Note that
the identification of galaxies in \citet{2007MNRAS.377....2M} was
based on a completely independent method, namely {\small SKID}
\citep{2001PhDT........21S}.  Note that {\small SKID} works best if
applied only on the distribution of star particles. Therefore, for the
sake of comparison, we excluded from our {\small SUBFIND} galaxies
those having no stellar component.

Figure~\ref{fig:comp_ss} shows a comparison for the {\it R3} run of
the positions, individual masses and stellar subhalo mass functions.
We find an excellent agreement between the stellar masses inferred
from both algorithms, in fact, the majority are identical for both
algorithms. Only for the very loosely bound star particles at the
periphery of the substructures some differences occur.  The middle
panel of the figure shows a comparison between the stellar masses of
the galaxies identified by the two algorithms. Note that the
horizontal and the vertical branches at low masses indicate the
galaxies which are identified by one of the two algorithms. However,
it is worth mentioning that there is no indication of any systematic
offset between the results, and most of the scatter arises because
sometimes merging substructures get split into two parts only by one
of the two algorithms.  In this respect, we find that {\small SUBFIND}
is marginally more efficient in separating two merging
substructures. As a general result of our comparison, we conclude that
our modifications of the {\small SUBFIND} algorithm are able to
reliably identify the stellar parts of subhalos even though we apply
the substructure finder to all the particles in the simulations.

\subsection{Resolution and convergence} \label{sec:numres}

To test our results we performed several DM-only runs by a changing a
number of parameters defining the numerical setup.  Among the
tests carried out, we mention the following: {\em (a)} use only the
Tree-based gravity solver and compare with the Tree-PM scheme; {\em
  (b)} increase the accuracy of time integration and force computation
far beyond the usual values assumed in our runs {\em (c)} change from
$z=5$ to $z=2$ the redshift at which the gravitational softening
switches from physical to comoving units.  In all these tests we
verified that the subhalo mass functions do not show appreciable
variations down to the mass corresponding to 20 self-bound DM
particles.

As already discussed in Section \ref{sec:sub_dm}, increasing
resolution in the DM runs produces numerically stable results in the
subhalo mass function, at least as long as only structures with at
least 20 gravitationally bound DM particles are selected. The
situation is expected to be different when considering instead
hydrodynamical simulations, which include physical processes that are
expected to be much more resolution dependent. The clearest example is
radiative cooling, whose efficiency is known to depend on the mass
resolution in cosmological simulations. In a similar fashion,
hydrodynamical processes, such as ram--pressure and viscous stripping,
depend quite sensitively on the resolution adopted.

In order to assess the effect of resolution on the subhalo mass
function, we carried out several simulations of a moderately poor
clusters, having a virial mass $M_{\rm vir}\simeq 2.3\times
10^{14}M_\odot$.  The set-up of the initial conditions of this cluster
have been described by \citet{2006MNRAS.367.1641B} (it corresponds to
the CL4 cluster of that paper). The cosmological model assumed in this
case is the same as that for the sample of clusters analysed in this
paper, except that a lower normalisation $\sigma_8=0.8$ of the power
spectrum was assumed \citep{2004MNRAS.348.1078B}.  Radiative
simulations of this cluster have been carried out by varying the
effective resolution in three different ways:
\begin{itemize}
\item Increase the mass resolution by a factor 4 and by a factor 15
  with respect to a reference resolution. The corresponding force
  resolutions are decreased by a factor $4^{1/3}$ and $15^{1/3}$,
  respectively (runs R1, R2 and R3 in Table \ref{tab:cpu}).
\item Change the criterion to assign gas particles in the initial
  conditions. Instead of using the same number of gas and DM
  particles, with a mass ratio reflecting the cosmic baryon fraction,
  we assigned one gas particle to every group of eight DM
  particles. In this way, the mass of each parent DM particle is
  decreased by an amount such to reproduce the cosmic baryon
  fraction. Given the standard values of the cosmic baryon fraction,
  this procedure gives comparable masses to DM and gas particles, thus
  reducing a possible spurious numerical heating of the gas particles
  induced by two--body encounters with DM particles. This scheme of
  adding gas to initial conditions can be seen as a way of increasing
  the accuracy of the gravity part (i.e.~increasing the number of DM
  particles) for a fixed resolution in the hydrodynamics. This
  prescription was originally suggested by \cite{1997MNRAS.288..545S}
  as a computationally efficient way of suppressing two--body heating
  in cooling flows.
\item Use different values for the minimum SPH kernel smoothing
  length, in units of the gravitational softening, $h_{\rm sm}$. The
  choice of this smoothing length is somewhat arbitrary. In radiative
  simulations, cooling causes gas to collapse and form high density
  cores at the centre of halos, thus causing the SPH smoothing length
  of cooled gas particles to fall below the gravitational
  softening. One therefore often restricts the SPH smoothing length to
  a certain fraction of the gravitational softening, ranging between
  zero (no restriction) and one (restricted to the gravitational
  softening). While the reference value adopted for our simulations is
  $h_{\rm sm}=0.1$, we have tested the effect of increasing it up to a
  value of unity.
\end{itemize}

To separate the effects of numerics from those related to the physics
included in the simulations, we decided to perform our study of
numerical stability for the simplest version of the radiative runs,
i.e.~by not using the kinetic feedback associated with galactic winds
({\em csfnw} runs).

\begin{table}
\begin{tabular}{|c|c|c|c|c|c|c|}
\hline
Run & $m_{\rm dm}$ & $m_{\rm gas}$ & $m_{*}$ & $M_{\rm sub}$ & $T_{\rm CPU}$ & System [GHz]\\
\hline
R1      & $2.2e9$ & $3.3e8$ & $1.7e8$ & $1e11$   & 55  & Opteron 2.8 \\
R2      & $6.6e8$ & $9.9e7$ & $4.9e7$ & $5e10$   & 173 & Opteron 2.8 \\
R3     & $1.5e8$ & $2.2e7$ & $1.1e7$ & $<1e10$  & 6126 & Power4 1.3 \\
R2-8   & $6.6e8$ & $7.9e8$ & $4.0e8$ & $7e10$   & 63   & Opteron 2.8 \\
R3-8   & $1.5e8$ & $1.8e8$ & $8.8e7$ & $2e10$   & 257  & Opteron 2.8 \\
\hline
\end{tabular}
\caption{The masses of the different particle species for the runs at
  different resolution. Column 1: run name. Columns 2-4: masses of the
  DM, gas and star particles, respectively (in units of
  $M_\odot$). Column 5: estimate of the stellar subhalo mass at which
  the simulations is numerically converged (since no higher resolution
  that R3 is achieved, we give only an upper limit in this
  case). Column 6: the CPU time (in hours) needed to perform the run.
} \label{tab:cpu}
\end{table}

\begin{figure*}
\includegraphics[width=0.33\textwidth]{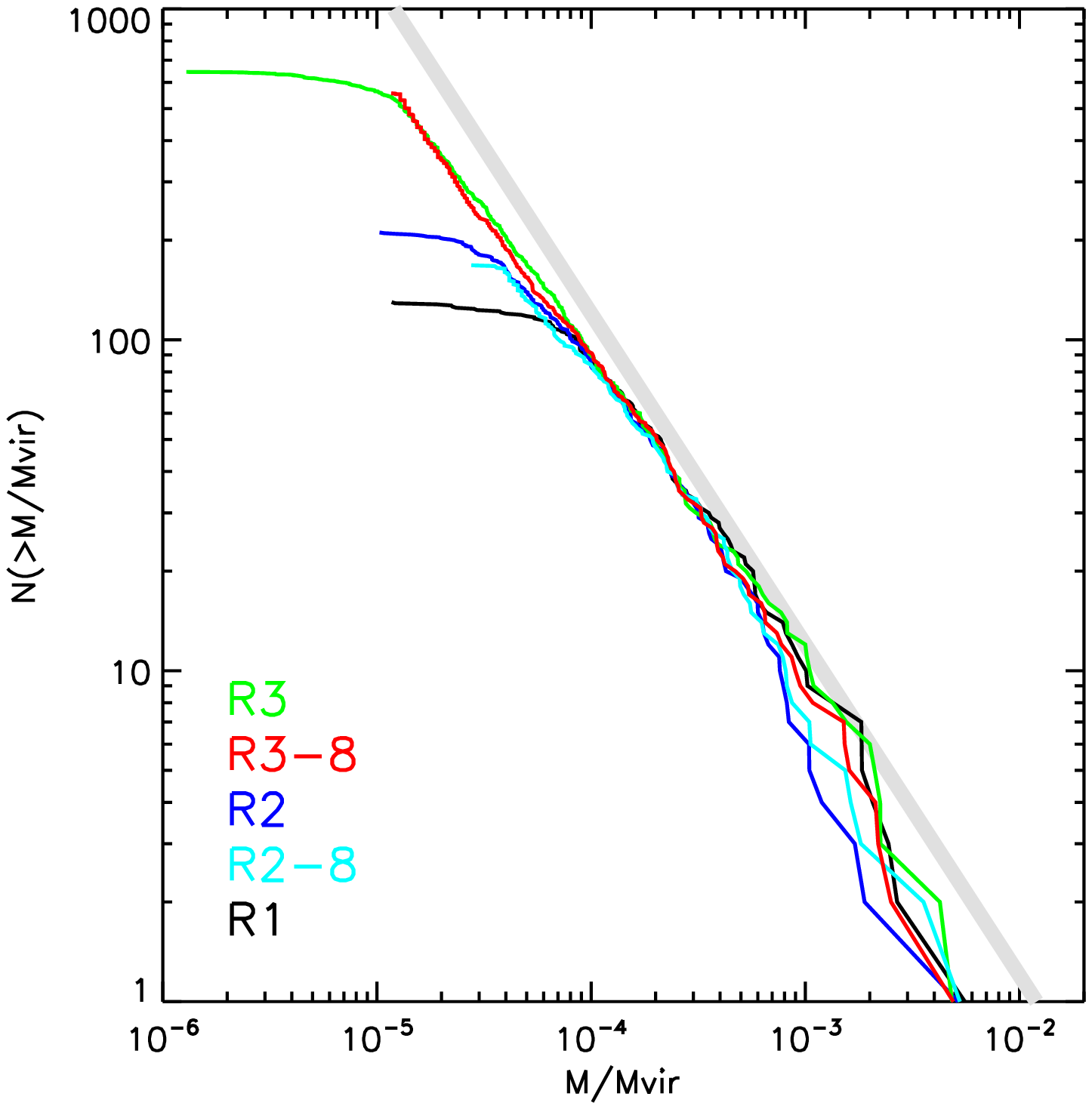}
\includegraphics[width=0.33\textwidth]{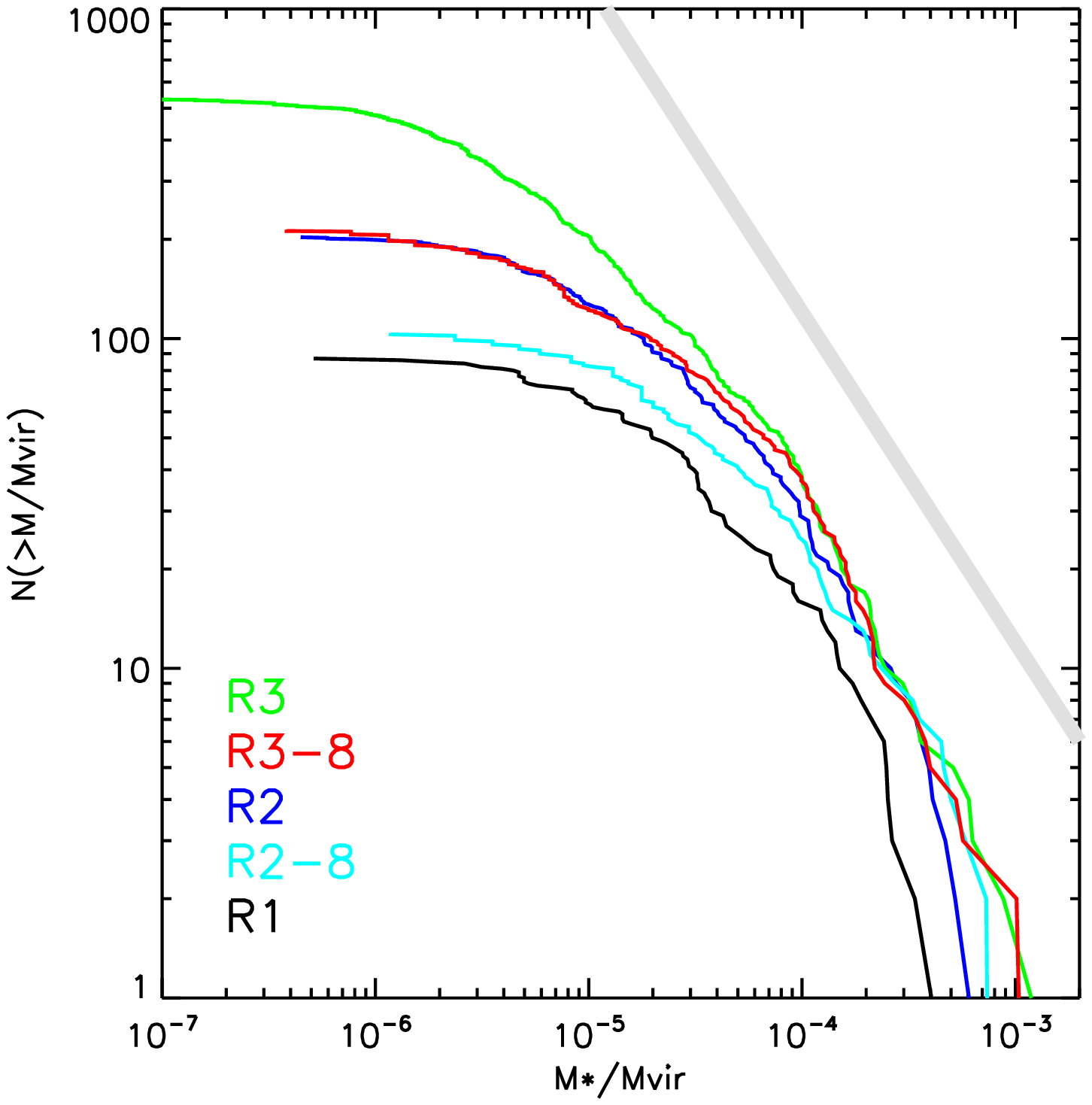}
\caption{Comparison of the total (left panel) and stellar (right panel)
  subhalo mass functions, for runs with increasing resolution as well as for
  runs with the alternative `gasification' scheme (see text for
  details). Whereas the total subhalo mass function seems to be converged for
  subhalos which are resolved by 30-50 dark matter particles, the stellar
  subhalo mass function appears to converge only much more slowly. Here the
  one over eight gasification scheme seems to work better, because it
  converges earlier for fewer gas particles and only mildly more dark matter
  particles.} \label{fig:fig7}
\end{figure*}

In the left panel of Figure~\ref{fig:fig7}, we show the results of our
resolution study.  They confirm that the mass range over which the
total subhalo mass function is reliably described widens from the R1
to the R3 run. In general, runs at varying resolution provide similar
results at masses corresponding to at least 20 DM particles, thus
confirming the result shown in Fig.~\ref{fig:fig_mf_dm} for DM-only
runs. Quite interestingly, the R2-8 and R3-8 runs, based on using
equal masses for DM and gas particles, provide results virtually
indistinguishable from the corresponding R2 and R3 runs over the whole
mass range where the results are stable against resolution. This
confirms that the total mass function is determined by the DM mass
resolution and that a numerically converged total mass function can be
obtained with the alternative way of ``gasifying'' initial conditions
at a lower computational cost (see Column 6 in Table~\ref{tab:cpu}).

The results are different for the stellar mass function, shown in the right
panel of Fig.~\ref{fig:fig7}. Since these simulations are not including an
effective feedback scheme able to regulate star formation, it is not
surprising that the subhalo mass function progressively increases without any
sign of convergence with resolution. A more interesting result concerns
instead the behaviour of the R2-8 and R3-8 runs. Despite the fact that these
runs have a poorer gas mass resolution than the R1 and R2 runs, respectively,
they produce a higher stellar subhalo mass function. For instance, the R3-8
run is consistent with the R3 run down to $1.5\times10^{10} M_\odot$, whereas
the R2 run converges on the R3 run only at $5\times10^{10} M_\odot$.  This
result is achieved with an acceptable increase of the computational cost. This
highlights that increasing the DM mass resolution is indeed an efficient way
of increasing the accuracy in the description of cooling in small subhalos,
without overly increasing the CPU time requirements.

\begin{figure*}
\includegraphics[width=0.33\textwidth]{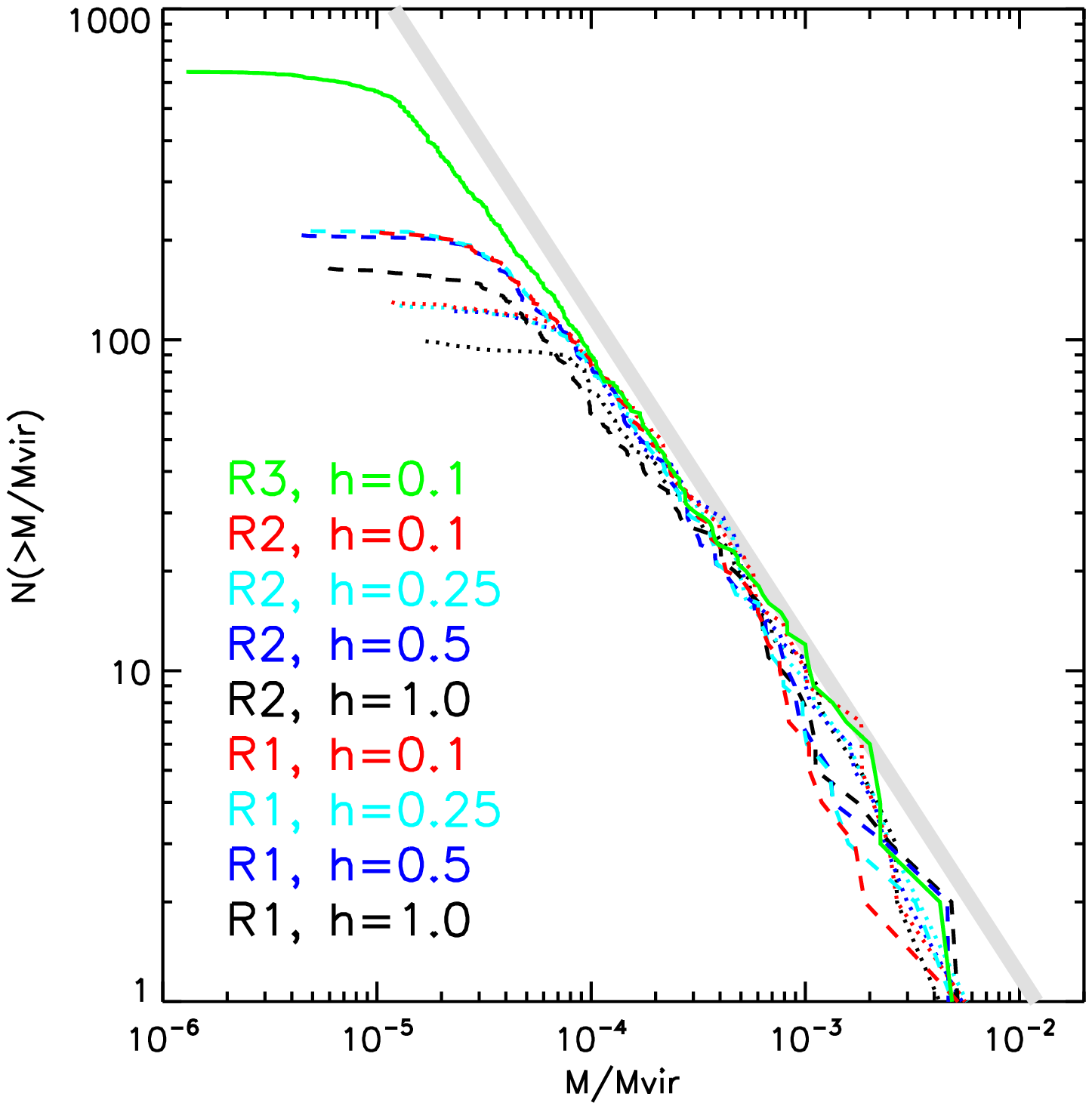}
\includegraphics[width=0.33\textwidth]{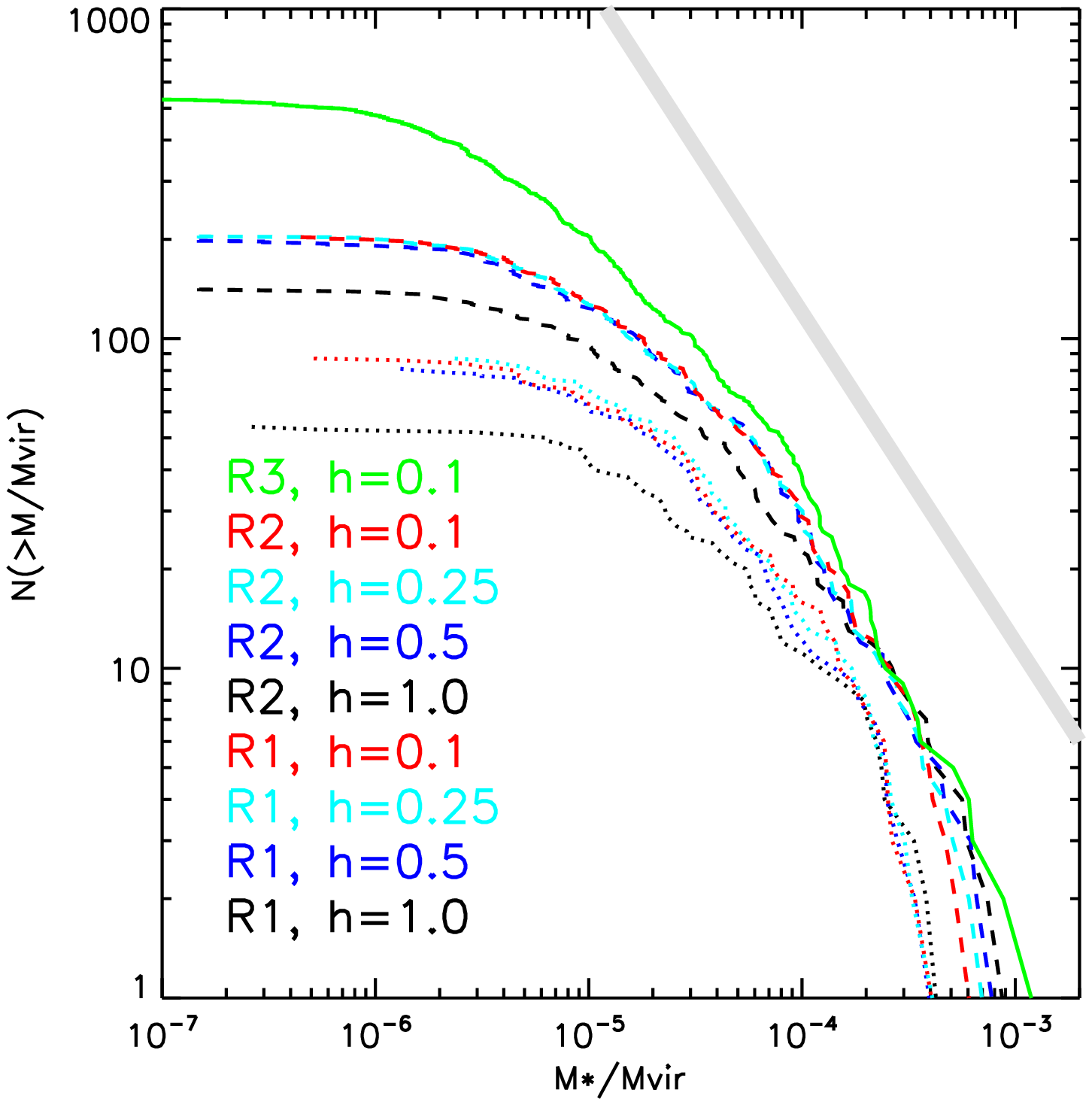}
\caption{The total (left panel) and stellar (right panel) subhalo
  mass functions at different resolutions and using, at each
  resolution, different values for the minimum of the SPH smoothing
  length, $h_{sm}$.} \label{fig:fig8}
\end{figure*}

Finally, Figure~\ref{fig:fig8} shows the effect of changing the minimum value
of the SPH smoothing length, $h_{\rm sm}$.  As for the total subhalo mass
function (right panel), it has only a mild effect at the smallest resolved
masses. We note that no differences exist for $h_{\rm sm}\le 0.5$, while a
significant reduction of the mass function takes place for $h_{\rm sm}=1$. The
effect is more apparent for the stellar mass function (left panel). This
result demonstrates that not allowing the SPH smoothing length to become
smaller than the gravitational softening makes the star clumps more fragile
within the strong tidal field of the cluster. We note that using a small
$h_{\rm sm}$ is computationally cheaper, since it prevents the occurrence of
computationally expensive SPH computations for a large number of neighbours in
the highly dense regions of cold gas.


\section{Database} \label{sec:database}

The results of our substructure analysis for the {\it dm}, {\it ovisc} and
{\it csf} version of all our simulated clusters (with the exception of {\it
  g696}) are publicly available within the German Virtual Observatory
(GAVO). The access to the data is realized by a VO-oriented, SQL-queryable
database similar to the one described in \citet{2006astro.ph..8019L}. The web
application that allows users to query the cluster database (called {\it
  Hutt}) is located at {\tt http://www.g-vo.org/HydroClusters}. Together with
this publication, the data structures are made available to all users. The
interface provides a full documentation of the available data and also gives
different examples for scientific {\tt SQL}-queries, returning data which can
be used to produce similar plots than presented in this paper. In the future,
we plan to allow users to register and to build up their own databases to
store intermediate results from the queries to allow more complex
analysis. Also, we plan to extend the underlying database with further
simulations that are not discussed in this study.

\end{document}